\documentclass[a4paper, 11pt, oneside]{article}
\pdfoutput=1
\usepackage{float}
\usepackage{jheppub}
\usepackage{epstopdf}
\usepackage{mathtools}
\usepackage{graphics}
\usepackage[normalem]{ulem}

\newcommand{\nc}{\newcommand}

\newlength{\absize}
\newcommand{\textoverline}[1]{$\overline{\mbox{#1}}$}
\nc{\non}{\nonumber}
\nc{\hc}{\hbox {H.c.}}
\nc{\noi}{\noindent}
\nc{\barx}{\bar{x}}
\nc{\pbarn}{\;\hbox {pb}}
\nc{\fbarn}{\;\hbox {fb}}
\nc{\lsp}{\;\;\;\;\;}
\nc{\Lsp}{\;\;\;\;\;\;\;\;\;\;}
\nc{\LLsp}{\lspace \lspace}
\nc{\lra}{\longrightarrow}
\nc{\beq}{\begin{equation}}  \nc{\eeq}{\end{equation}}
\nc{\bea}{\begin{eqnarray}}  \nc{\eea}{\end{eqnarray}}
\nc{\baa}{\begin{array}}     \nc{\eaa}{\end{array}}
\nc{\bit}{\begin{itemize}}   \nc{\eit}{\end{itemize}}
\nc{\ben}{\begin{enumerate}} \nc{\een}{\end{enumerate}}
\nc{\bce}{\begin{center}}    \nc{\ece}{\end{center}}
\nc{\bpm}{\begin{pmatrix}}   \nc{\epm}{\end{pmatrix}}
\nc{\bvt}{\begin{verbatim}}  \nc{\evt}{\end{verbatim}}
\def\lsim{\mathrel{\raise.3ex\hbox{$<$\kern-.75em\lower1ex\hbox{$\sim$}}}}
\def\gsim{\mathrel{\raise.3ex\hbox{$>$\kern-.75em\lower1ex\hbox{$\sim$}}}}

\def\lcal{{\cal L}}
\def\mcal{{\cal M}}

\def\pcal{{\cal P}}

\nc{\tanb}{\tan\beta}
\nc{\mch}{M_{H^\pm}}

\def\thetaW{{\theta}_{\rm W}}

\def\mch{M_{H^\pm}}
\def\mchsq{M_{H^\pm}^2}

\nc{\for}{\lsp {\rm for} \lsp}
\nc{\andd}{\lsp {\rm and} \lsp}

\renewcommand{\Re}{\mbox{Re\thinspace}}
\renewcommand{\Im}{\mbox{Im\thinspace}}
\newcommand{\half}{{\textstyle\frac{1}{2}}}

\allowdisplaybreaks 

\renewcommand{\vec}[1]{\boldsymbol{#1}}

\title{Symmetries of the 2HDM: an invariant formulation and consequences}
\author[a,b]{P.M. Ferreira,}
\affiliation[a]{Instituto Superior de Engenharia de Lisboa~---~ISEL, 1959-007~Lisboa, Portugal}
\affiliation[b]{Centro de F\'isica Te\'orica e Computacional, Faculdade de Ci\^encias,
Universidade de Lisboa, Campo Grande, 1749-016~Lisboa, Portugal}
\author[c]{B. Grzadkowski,}
\affiliation[c]{Faculty of Physics, University of Warsaw, Pasteura 5, 02-093 Warsaw, Poland}
\author[d]{O. M. Ogreid,}
\affiliation[d]{Western Norway University of Applied Sciences,
Postboks 7030, N-5020 Bergen, Norway, }
\author[e]{P. Osland}
\affiliation[e]{Department of Physics and Technology,
University of Bergen, Postboks 7803, N-5020 Bergen, Norway}
\emailAdd{pmmferreira@fc.ul.pt}
\emailAdd{bohdan.grzadkowski@fuw.edu.pl}
\emailAdd{omo@hvl.no}
\emailAdd{Per.Osland@uib.no}
\date{\today}

\abstract{Symmetries of the Two-Higgs-Doublet Model (2HDM) potential that can be extended to the whole Lagrangian, i.e. the CP-symmetries CP1, CP2, CP3 and the Higgs-family symmetries $Z_2$, U(1) and SO(3) are discussed. Sufficient and necessary conditions
in terms of constraints on masses and physical couplings for the potential to respect each of these symmetries are found.
Each symmetry can be realized through several alternative cases, each case being a set of relations among physical parameters.
We will show that some of those relations are invariant under the renormalization group, but others are not.
The cases corresponding to each symmetry group are illustrated by analyzing the interplay between the
potential and the vacuum expectation values.}


\begin{document}

\maketitle

\flushbottom

\section{Introduction}
\label{sect:Introduction}
\setcounter{equation}{0}

The discovery of the Higgs boson by the LHC collaborations in 2012~~\cite{Aad:2012tfa,Chatrchyan:2012xdj}
was a remarkable achievement. A series of precision measurements of its properties
(see for instance~\cite{Aad:2015zhl,Khachatryan:2016vau}) revealed that the particle observed
at the LHC has spin 0 and behaves, to a good degree of precision, as one would expect for the
Higgs boson in the Standard Model (SM) \cite{Yang:1954ek,Glashow:1961tr,Weinberg:1967tq,Salam:1968rm,Higgs:1964pj,Englert:1964et}. Since then, a great deal of effort has been put into searches for
Beyond the Standard Model (BSM) physics, but so far no significant deviation from SM phenomenology
has been observed, no significant excess hinting at new particle resonances has been found.
The coming years will bring about a wealth of LHC results as we progress towards its high
luminosity phase. This will provide us with the opportunity to further test BSM theories, some
of which are already quite constrained by current data. The drive to extend the Standard Model
is obvious when one considers the amount of observed facts that the model does {\it not} explain:
the hierarchy in fermion masses, the astrophysical and cosmological data indicating the existence of
Dark Matter, and the universe's matter--antimatter asymmetry, among other puzzles.

There are many interesting proposals for BSM physics. One of the most popular consists in enlarging
the scalar sector, and one of the simplest models of this kind is the Two-Higgs-Doublet Model
(2HDM), proposed in 1973 by Lee~\cite{Lee:1973iz} as a means to obtain an extra source of CP violation
from spontaneous symmetry breaking
(see~\cite{Gunion:1989we,Branco:2011iw}). In this model the SM field content is complemented with a second
SU(2) doublet, which yields a larger scalar spectrum -- a charged scalar field and three neutral ones
(in versions of the model where CP is conserved two of those scalars are CP-even, the third being odd).
The model has a rich phenomenology, and different versions of the 2HDM allow for dark matter candidates, spontaneous or explicit CP violation, tree-level flavour-changing neutral currents (FCNC) mediated by scalars, and many other interesting phenomena. In fact, these ``versions"
of the 2HDM correspond in many cases to different {\it symmetries} imposed on the model, which reduce the
number of free parameters (thus increasing its predictive power) and change the phenomenology of
the theory. The first such symmetry was introduced by Glashow, Weinberg and Paschos~\cite{Glashow:1976nt,
Paschos:1976ay} -- a discrete $Z_2$ symmetry corresponding to one of the doublets being odd under it, which, when
extended to the whole Lagrangian, eliminated the tree-level FCNC mentioned above. Another symmetry,
a continuous U(1), was proposed by Fayet \cite{Fayet:1974fj}, in a model linking spontaneous P violation to interchange of the two doublets, as well as in the context of R-symmetry \cite{Fayet:1974pd} and other SUSY applications \cite{Fayet:1975yi,Fayet:1987js}. The U(1) symmetry was also invoked by Peccei and Quinn~\cite{Peccei:1977hh} in an attempt to solve
the strong CP problem. U(1) and $Z_2$ symmetries are also relevant for models of 2HDM-based cosmic strings and magnetic monopoles \cite{Eto:2018tnk,Eto:2019hhf,Eto:2020hjb,Eto:2020opf}.
Other symmetries were also proposed and thoroughly studied.

The study of 2HDM symmetries, however, is complicated by the fact that the model possesses
a {\it basis invariance}. In fact, a general 2HDM can be formulated
adopting different bases for the doublets, therefore, e.g., the scalar sector of the model
is not uniquely defined. Different (while being physically equivalent) potentials could be related
by a U(2) basis transformation which is not a symmetry of the model. Such (basis) transformations
will in general make the parameters of the potential change, and a symmetry of the potential that is
 manifest in one basis will in general not be obvious in another.
Therefore, the symmetries might be hidden and difficult to recognize.
However, there exist physical parameters of the scalar sector of the 2HDM that are independent of the
basis adopted to formulate the model and those could be utilized in the identification of the symmetries.
The ultimate goal of this work is to provide a formulation of all possible symmetries of the 2HDM
potential in terms of physical (observable)
parameters, like masses and measurable coupling constants. Knowing the physical symmetry conditions
would make the verification of invariance unambiguous, without any reference to a particular
basis.

Symmetries are of fundamental relevance both for classical and for quantum field theories. Hereafter we limit
ourselves to internal symmetries, even though
space-time transformations also play a fundamental role in contemporary physics, the Lorentz invariance in
special relativity and reparametrization invariance in general relativity being famous examples. The presence
of continuous global symmetries implies, via the Noether theorem, the existence of conserved currents and
charges. Conservation of electric charge,
lepton or baryon numbers could serve as other examples of consequences of U(1) invariance. Even if a
symmetry is broken explicitly by
the presence of non-invariant terms in the Lagrangian, still, if the breaking is {\it small}, the notion
 of the symmetry might still be very useful.
When global continuous symmetries are not respected by vacuum states the Goldstone theorem
requires the existence of massless scalars that correspond to all the broken generators of the symmetry
group. Here again the role of the symmetry is crucial while trying to understand the mass spectrum of
particles, pions as the Goldstone bosons of spontaneously broken $\text{SU(2)}_L \times \text{SU(2)}_R$ serve here as
a spectacular example. On the other hand, if a symmetry remains unbroken, its presence implies constraints
on parameters,
e.g. mass degeneracies appear and/or some couplings are related, while others might vanish.
This is why symmetries are one of the main tools for model building.
If continuous symmetries are local, their importance is even amplified, as in that case they lead to
gauge theories such as QED or the SM itself.
When symmetries are broken by terms of dimension~$<4$ (``soft'' symmetry breaking terms) then, according to Symanzik
 \cite{Symanzik:1969ek,Symanzik:1970zz}, a theory remains renormalizable. That is yet another illustration
 of the power of symmetries.

Another class of symmetries is formed by discrete transformations that leave the action invariant.
In particular, space (P) and time (T) reflections, and  charge conjugation (C) have to be emphasized.
In fact, composite symmetries such as CP and CPT
play fundamental roles in quantum field theory. Discrete symmetries are also often adopted in theories
of dark matter where e.g.\ the $Z_2$ symmetry mentioned above may be used
to stabilize DM particles (taken to be odd under the symmetry).
One may then wonder: {\it how many different internal symmetries can one impose on the 2HDM scalar sector}?
The answer was found by Ivanov \cite{Ivanov:2006yq,Ivanov:2007de}, using a bilinear field formalism
to prove that there were only {\it six} different classes of symmetries that could be imposed. Three of these
were so-called {\it Higgs family symmetries}, in which invariance of the scalar potential
is required for doublet transformations of the form $\Phi_i \rightarrow \Phi_i^\prime = U_{ij}\Phi_j$,
 with $U_{ij}$ elements of a $2\times 2$ unitary matrix $U$, they include the $Z_2$ symmetry, for which
$U = $ diag$(1,-1)$, and the Peccei--Quinn U(1) symmetry, $U = $ diag$(1,e^{i\theta})$ for a
generic real phase $\theta$, mentioned above; and the SO(3) symmetry, for which one takes a
general U(2) matrix. The remaining three symmetries arise from requiring invariance under
{\it generalised CP transformations} of the form $\Phi_i \rightarrow \Phi_i^\prime = X_{ij}\Phi_j^*$,
where again $X$ is a unitary $2\times 2$ matrix. Different choices of $X$ yield different CP symmetries,
to wit CP1 (the ``standard" CP symmetry, with $X$ equal to the identity), a discrete CP2 symmetry
\cite{Davidson:2005cw,Maniatis:2007vn,Ferreira:2009wh}
and a continuous CP3 one~\cite{Ferreira:2010bm}. These, then, are the only six symmetries for an
$\text{SU(2)}\times \text{U(1)}$ invariant 2HDM scalar potential. If one chooses to ignore hypercharge, then other symmetries arise, such as the custodial symmetry. The full classification of
those possibilities, which we will not consider in the present work, may be found in
\cite{Battye:2011jj,Pilaftsis:2011ed}.

We are going to find basis-independent conditions for invariance of the 2HDM potential under the field
 transformations which yield the six 2HDM symmetry classes mentioned above. Such conditions were
 expressed in a covariant way in terms of basis-dependent parameters in Ref.~\cite{Ferreira:2010yh},
 whereas here we shall express these conditions in terms of basis-independent observables, and discuss
 spontaneous breaking of the symmetries.
This work is a natural extension of the papers \cite{Grzadkowski:2014ada,Grzadkowski:2016szj},
where we have discussed the invariant formulation of the 2HDM under CP transformation.
Here, our intention is to provide conditions (in terms of measurable parameters)
for invariance of the 2HDM potential under all the remaining possible symmetries.

The symmetries in question satisfy the following hierarchy \cite{Ferreira:2010yh},
\begin{equation}
CP1 \subset Z_2 \subset \left\{U(1)\atop CP2 \right\} \subset CP3 \subset SO(3).
\end{equation}
These relations will be reflected by the physical constraints to be quoted in the following.

The formulation of basis-independent conditions for global symmetries in the 2HDM has recently been addressed by Bento {\em et al} \cite{Bento:2020jei} in the framework of a rather mathematical formalism. While that approach is general and may have interesting applications also in other theories, we think that at least in the case of 2HDM our approach is more useful, directly expressing constraints in terms of physical quantities.

The paper is organized as follows. In section~\ref{sect:The model} we review the model, and discuss the choice of parameters.
In section~\ref{Sec:Preliminaries} we review the approach of Ref.~\cite{Ferreira:2010yh}, and outline the mapping to physical parameters.
Then, in section~\ref{sect:Results} we present our results for the different cases in compact form, with the detailed analysis presented in section~\ref{Sec:analysis}. In section~\ref{sec:rge} we address the issue of stability under the renormalization group equations (RGE),
and in section \ref{Sec:summary} we provide a brief discussion, highlighting the RGE-stable cases.
More technical material is collected in three appendices.

\section{The model}
\label{sect:The model}
\setcounter{equation}{0}

We shall start out by parametrizing the scalar potential of the generic (CP-violating) 2HDM in the common fashion:\footnote{We shall use the same notation, $\Phi$ and $\lambda$, both in a generic basis and in a Higgs basis. It will be clear from the context which basis is adopted.}
\begin{align}
V(\Phi_1,\Phi_2) &= -\frac12\left\{m_{11}^2\Phi_1^\dagger\Phi_1
+ m_{22}^2\Phi_2^\dagger\Phi_2 + \left[m_{12}^2 \Phi_1^\dagger \Phi_2
+ \hc\right]\right\} \nonumber \\
& + \frac{\lambda_1}{2}(\Phi_1^\dagger\Phi_1)^2
+ \frac{\lambda_2}{2}(\Phi_2^\dagger\Phi_2)^2
+ \lambda_3(\Phi_1^\dagger\Phi_1)(\Phi_2^\dagger\Phi_2) \nonumber \\
&+ \lambda_4(\Phi_1^\dagger\Phi_2)(\Phi_2^\dagger\Phi_1)
+ \frac12\left[\lambda_5(\Phi_1^\dagger\Phi_2)^2 + \hc\right]\nonumber\\
&+\left\{\left[\lambda_6(\Phi_1^\dagger\Phi_1)+\lambda_7
(\Phi_2^\dagger\Phi_2)\right](\Phi_1^\dagger\Phi_2)
+{\rm \hc}\right\}.
\label{Eq:pot}
\end{align}
All parameters in (\ref{Eq:pot}) are real, except for $m_{12}^2$, $\lambda_5$, $\lambda_6$ and $\lambda_7$,
which in general could be complex.

\subsection{Choice of basis and basis independence}
\label{sect:Choice_of_basis}
The potential has been written out in terms of two doublets that we have named $\Phi_1$ and $\Phi_2$.
Since both doublets have identical quantum numbers and there is nothing {\it a priori} to distinguish
them, we could equally well have expressed the potential in terms of linear combinations of these (initial)
 doublets, i.e. if we define $\bar\Phi_i=U_{ij}\Phi_j$, where $U$ is a U(2)-matrix, we can instead choose to
 express the potential in terms of $\bar\Phi_1$ and $\bar\Phi_2$. This is referred to as a change of
 basis. Note that the parameters of the potential will in general change under a change of basis. How
 the parameters change under the most general change of basis is given explicitly in Eqs. (5)--(15) of
 \cite{Gunion:2005ja}.
This reparametrization freedom  means that some of the parameters in (\ref{Eq:pot}) are superfluous and
can be eliminated via a judicious basis choice. Thus, the number of free parameters in the most general 2HDM
potential is not the 14 shown in (\ref{Eq:pot}), but rather 11 \cite{Davidson:2005cw}, as we will discuss
later on. But basis changes also introduce complications when it comes to an attempt to recognize whether
a given 2HDM potential is invariant under a particular symmetry.

Clearly, physics cannot depend on an arbitrary choice of basis for the Higgs doublets. All measurable quantities must be basis independent, thereby leading to the study of basis invariant quantities in multi-Higgs-Doublet Models (NHDMs).  Of course,  the scalar masses are basis invariant. The same holds for most of the physical couplings, the exception being the couplings $f_i$ that are defined in section~\ref{sect:Scalar_couplings}. They occur in couplings involving charged fields, whose phases are arbitrary, and are thus pseudo-invariants~\cite{Grzadkowski:2014ada}.

In the present work we shall derive relations between the basis-invariant masses and couplings needed in order to respect certain symmetries imposed upon the potential. In order to do so we shall choose to derive these relations in a particularly simple basis, namely the Higgs basis \cite{Donoghue:1978cj,Georgi:1978ri}. The Higgs basis is a basis in which only one doublet has a non-vanishing, real and positive vacuum expectation value (VEV), whereas the other doublet has a vanishing VEV.  If the original doublets have neutral
(and in general complex) VEVs
\begin{equation}
\langle \Phi_1 \rangle=\frac{1}{\sqrt{2}}\left(
\begin{array}{c}0\\
v_1
\end{array}\right),\quad
\langle \Phi_2 \rangle =\left(
\begin{array}{c}0\\
v_2
\end{array}\right),
\end{equation}
then the Higgs basis is obtained via the field redefinition\footnote{The Higgs basis is not unique, it is still possible to perform a basis change consisting of a U(1) rotation on $\Phi_2$, staying within the Higgs basis. Furthermore, we will omit the bar from the doublets when working in the Higgs basis. It will be clear from the context in which basis we choose to work.}
\begin{equation} \label{Hbasis}
\bar\Phi_1\equiv \frac{v^*_1 \Phi_1+v^*_2\Phi_2}{v}\,,
\qquad\quad \bar\Phi_2 \equiv\frac{-v_2 \Phi_1+v_1\Phi_2}{v}
 \,,
\end{equation}
so that the new fields have VEVs given by
\begin{equation}
\langle \bar\Phi_1 \rangle=\frac{1}{\sqrt{2}}\left(
\begin{array}{c}0\\
v
\end{array}\right),\quad
\langle \bar\Phi_2 \rangle =\left(
\begin{array}{c}0\\
0
\end{array}\right),
\end{equation}
with $v=\sqrt{v_1^2+v_2^2}=246\, {\rm GeV}$. We must make sure that the vacuum corresponds to a minimum of the potential, and by demanding that the derivatives of the potential with respect to the fields should vanish we end up with the stationary-point equations in the Higgs basis,
\begin{equation}
m_{11}^2=v^2\lambda_1,\quad
m_{12}^2=v^2\lambda_6.
\end{equation}

Demanding that the vacuum should correspond to a stationary point does not guarantee that it is a minimum of the potential. One must also demand that the squared masses of the physical scalars are positive in order for the potential to have the curvature of a minimum point. In the present study we shall encounter situations where some physical scalar has a vanishing mass. Then we shall relax the requirement of positive squared masses by simply demanding that the physical scalars have non-negative squared masses. One should also add that within
the 2HDM there may be coexisting minima for the same set of parameters \cite{Ivanov:2006yq,Ivanov:2007de,Barroso:2007rr,Barroso:2013ica,Barroso:2013awa,Ivanov:2015nea},
so in fact one must also verify whether the minimum we are interested in is the global one. We will however not address this issue in the present work.

\subsection{Scalar fields and mass eigenstates}
\label{sect:Parametrization_of_doublets}
Having chosen to work within the Higgs basis, we may parametrize the two doublets as
\begin{equation}
\Phi_1=\left(
\begin{array}{c}G^+\\ (v+\eta_1+iG^0)/\sqrt{2}
\end{array}\right), \quad
\Phi_2=\left(
\begin{array}{c}H^+\\ (\eta_2+i\eta_3)/\sqrt{2}
\end{array}\right).
\end{equation}
The great advantage of working in the Higgs basis is that the massless Goldstone fields,
which we represent here by $G^0$ and $G^\pm$,
are immediately present in the VEV-carrying doublet. Then,
$H^\pm$ are the massive charged scalars. The neutral fields $\eta_i$ are not mass eigenstates, so we relate them to the mass eigenstate fields $H_i$ (whose CP properties are in general undefined)
by an orthogonal rotation matrix $R$ as
\begin{equation} \label{Eq:R-def}
\begin{pmatrix}
H_1 \\ H_2 \\ H_3
\end{pmatrix}
=R
\begin{pmatrix}
\eta_1 \\ \eta_2 \\ \eta_3
\end{pmatrix}.
\end{equation}
As for the charged sector, the masses of the charged scalars $H^\pm$ can be read directly off from the corresponding bilinear terms in the potential, and are given in the Higgs basis by
\begin{eqnarray}
M_{H^\pm}^2=-\frac{m_{22}^2}{2}+\frac{v^2}{2} \lambda_3.\label{chargedmass}
\end{eqnarray}
As for the neutral sector, the bilinear terms can be written as
\begin{equation}
\frac{1}{2}
\left(
\begin{array}{ccc}
\eta_1 & \eta_2 & \eta_3
\end{array}
\right)
\mcal^2
\left(
\begin{array}{c}
\eta_1 \\ \eta_2 \\ \eta_3
\end{array}
\right),
\end{equation}
where the mass-squared matrix is in the Higgs basis found to be
\begin{eqnarray}
&&{\cal M}^2
=v^2
\begin{pmatrix}
\lambda_1 & \Re \lambda_6 & -\Im \lambda_6 \\
\Re \lambda_6 & \frac{1}{2}(\lambda_3+\lambda_4+\Re \lambda_5-\frac{m_{22}^2}{v^2}) & -\frac{1}{2}\Im \lambda_5 \\
-\Im \lambda_6
& -\frac{1}{2}\Im \lambda_5 & \frac{1}{2}(\lambda_3+\lambda_4-\Re \lambda_5-\frac{m_{22}^2}{v^2})
\end{pmatrix}.\label{neutralmassmatrix}
\end{eqnarray}
Then, by using (\ref{Eq:R-def}) we obtain the masses of the neutral scalars
from the diagonalization
of the mass-squared matrix, $\mcal^2$,
\begin{equation}
\label{Eq:cal-M}
{\rm diag}(M_1^2,M_2^2,M_3^2)=R{\cal M}^2R^{\rm T}.
\end{equation}
We shall use indices $i,j,k\in\{1,2,3\}$ to refer to these neutral mass eigenstates.

\subsection{Physical couplings of scalar eigenstates}
\label{sect:Scalar_couplings}
Having identified and diagonalized the mass terms of the potential, the remaining terms are trilinear and quadrilinear in the scalar fields, thereby representing trilinear and quadrilinear couplings among the scalars. Some of these couplings play an important role in the present work, namely the three trilinear neutral--charged $H_iH^+H^-$ couplings and the quartic charged self-interaction, that is $H^+H^+H^-H^-$. We denote these by $q_i$ and $q$, respectively. In the Higgs basis they are given by
\begin{eqnarray}
q_{i}&\equiv&{\rm Coefficient}(V,H_iH^+H^-)\label{eq:qi}
\nonumber\\
&=&v(R_{i1}\lambda_3+R_{i2}\Re \lambda_7-R_{i3}\Im \lambda_7),\\
q&\equiv&{\rm Coefficient}(V,H^+H^+H^-H^-)\label{eq:q}
\nonumber\\
&=&\frac{1}{2}\lambda_2.
\end{eqnarray}
One can show explicitly that these couplings are all basis independent \cite{Ogreid:2018bjq}. The LHC is already probing one of these couplings, $q_1$, via the diphoton decay of the discovered Higgs boson, since in the 2HDM a scalar loop contributes to that amplitude.
The scalar--gauge boson couplings will also be necessary for the present work. They
originate from the kinetic term of the Lagrangian, which may be written as
\begin{equation} \label{Eq:gauge-IS}
\lcal_k=(D_\mu \Phi_1)^\dagger(D^\mu \Phi_1) + (D_\mu \Phi_2)^\dagger(D^\mu \Phi_2),
\end{equation}
where we have adopted the usual definitions,
 $D^\mu=\partial^\mu+\frac{ig}{2}\sigma_iW_i^\mu+i\frac{g^\prime}{2}B^\mu$,
$W_1^\mu=\frac{1}{\sqrt{2}}(W^{+\mu}+W^{-\mu})$,
$W_2^\mu=\frac{i}{\sqrt{2}}(W^{+\mu}-W^{-\mu})$,
$W_3^\mu=\cos\thetaW Z^\mu+\sin\thetaW A^\mu$ and
$B^\mu=-\sin\thetaW Z^\mu+\cos\thetaW A^\mu$.
Relevant couplings can now be read off from the kinetic terms,
\bea
{\rm Coefficient}\left(\lcal_k,Z^\mu \left[H_j \overleftrightarrow{\partial_\mu} H_i\right]\right)&=&\frac{g}{2v\cos\thetaW}\epsilon_{ijk}e_k,\\
{\rm Coefficient}\left(\lcal_k,H_i Z^\mu Z^\nu\right)&=&\frac{g^2}{4\cos^2\thetaW}e_i\,g_{\mu\nu},\\
{\rm Coefficient}\left(\lcal_k,H_i W^{+\mu} W^{-\nu}\right)&=&\frac{g^2}{2}e_i\,g_{\mu\nu},\\
{\rm Coefficient}\left(\lcal_k,(H^+\overleftrightarrow{\partial_\mu} H_i )W^{-\mu}\right)&=&i\frac{g}{2v}f_i,\\
{\rm Coefficient}\left(\lcal_k,(H^-\overleftrightarrow{\partial_\mu} H_i )W^{+\mu}\right)&=&-i\frac{g}{2v}f^*_i.
\eea
It is not a coincidence that different vertices are proportional
to the same quantities $e_i$ and $f_i$, but rather a consequence of the gauge invariance of the model.
The factors $e_i$ and $f_i$ are given, in terms of Higgs basis parameters, by
\begin{equation}
e_i\equiv vR_{i1},\quad
f_i\equiv v\left(R_{i2}-iR_{i3}\right).
\end{equation}
In a general basis, the factors $e_i=v_1R_{i1}+v_2R_{i2}$ are found to be explicitly invariant under a
change of basis \cite{Ogreid:2018bjq}.
Unitarity of the rotation matrix in this multi-doublet model forces these
factors to  satisfy a sum rule, to wit
\bea
e_1^2+e_2^2+e_3^3=v^2.
\eea
The factors $f_i$ (and their conjugate partners $f_i^*$) appear in couplings between scalars and gauge bosons whenever an $H^+W^-$ pair ($H^-W^+$ pair) is present at the vertex. In a general basis they are given by $f_i= v_1R_{i2}-v_2R_{i1}-ivR_{i3}$. These factors are not invariant under a change of basis, they transform as pseudo-invariants, meaning that their lengths are invariant, but their phases change, see \cite{Grzadkowski:2018ohf}. The product $f_if_j^*$ is, however, invariant under a change of basis. This can also be seen from the following identity
\bea
f_if_j^*&=&v^2\delta_{ij}-e_ie_j+iv\epsilon_{ijk}e_k.
\label{fifj}
\eea

\subsection{The physical parameter set}
\label{sect:Physical_parameters}
While the potential of the 2HDM has a total of 14 real parameters, the number of observable quantities arising from the potential is in fact less than 14. Through a series of basis changes one can reduce the number of potential parameters from 14 to 11, leaving us with a total of 11 physical independent quantities as stated in \cite{Davidson:2005cw}.
A simple way of seeing this is by considering once again the most general 2HDM potential of
\eqref{Eq:pot} -- it is easy to imagine a doublet rotation such that $m_{12}^2$ is set to zero, thus
eliminating two parameters from the potential (since this coefficient is in general complex). With this
``diagonalization" of the quadratic part of the potential the quartic couplings will also change, of course. Then, with $m_{12}^2 = 0$ in the new basis we can still rephase one of the (new) doublets to absorb a complex phase from $\lambda_5$, for example, thus eliminating a third parameter.

It is in principle possible to devise experiments from which one can make 11 independent measurements of quantities arising from the bosonic sector of the 2HDM, and from these mesurements one can reconstruct the parameters of the potential.
Here, instead of working with 11 independent potential parameters, we will choose a set of 11
parameters consisting of masses and bosonic couplings, that we denote by $\pcal$ \cite{Grzadkowski:2014ada,Grzadkowski:2016szj,Grzadkowski:2018ohf,Ogreid:2018bjq}. For this purpose we pick the mass of the charged scalars as well as the masses of the three neutral scalars along with the scalar couplings  $H_iH^+H^-$ and $H^+H^+H^-H^-$ and the coefficients $e_i$ of the gauge couplings  to get\footnote{Note that the couplings $e_i$ and $q_i$ have dimension of mass.}
\begin{equation} \label{Eq:pcal}
{\cal P}\equiv\{M_{H^\pm}^2,M_1^2,M_2^2,M_3^2,e_1,e_2,e_3,q_1,q_2,q_3,q\},
\end{equation}
which we denote as our physical\footnote{Masses and couplings, being basis invariant, are more closely related to what one can measure in experiments than the potential parameters, which may be basis dependent. Therefore we have chosen to call $\pcal$ the ``physical'' parameter set. Basis independence is, however, only a necessary requirement -- not a sufficient condition for a quantity to be measurable. In fact, we shall encounter some situations where some $e_i$ and $q_i$ lose their physical meaning. This happens for some models with mass degeneracy at tree level, where the degeneracy is lost at loop level. Details are given in Chapters \ref{Sec:analysis} and \ref{sec:rge}. Bearing this in mind, we continue to use the name ``physical'' for the parameter set $\pcal$.} parameter set, consisting of 11 independent invariant quantities. All the other purely scalar couplings of the model are expressible in terms of these 11 parameters \cite{Grzadkowski:2018ohf} along with the auxiliary complex couplings $f_i$ and $f_i^*$ (which do not appear separately in physical observables because of (\ref{fifj})). All physical properties of the scalar sector are thus expressible in terms of masses and couplings.
{\it For a 2HDM with some symmetry, then, some of the 11 parameters of the physical parameter set $\pcal$ will either be related or set to zero.}

\section{The bilinear formalism and symmetries}
\label{Sec:Preliminaries}
\setcounter{equation}{0}
In this section we will briefly review the bilinear formalism, in which the scalar potential
is expressed not in terms of the doublets themselves but rather using their gauge-invariant bilinear
products. This formalism is rather useful when studying symmetries and possible vacua of NHDM models.
An earlier application of this method appeared in~\cite{Velhinho:1994np} and was used to establish
tree-level theorems about the stability of 2HDM minima \cite{Ferreira:2004yd,Barroso:2005sm,Barroso:2007rr}.
A remarkable formulation of bilinears in a Minkowski space was developed
in~\cite{Nishi:2006tg,Ivanov:2006yq,Ivanov:2007de,Nishi:2007nh,Nishi:2007dv}. The bilinear formulation
used in this paper is that
of~\cite{Maniatis:2007vn,Maniatis:2006fs,Maniatis:2007de,
Maniatis:2009vp,Ferreira:2010hy}. The formalism was adopted to investigate the custodial symmetry of the 2HDM in~\cite{Grzadkowski:2010dj}.
Similar formalisms have also been used for other models, for instance
the 3HDM~\cite{Ivanov:2010ww,Ivanov:2014doa}, the complex singlet--doublet model
\cite{Ferreira:2016tcu} and the N2HDM~\cite{Ferreira:2019iqb,Engeln:2020fld}.

\subsection{Field bilinears}
\label{Sec:Bilinear notation}
It is very convenient to express the potential in terms of four gauge-invariant bilinear
products of the doublets. We will follow closely the conventions of \cite{Ferreira:2010yh},
defining the bilinears as
\begin{alignat}{2}
K_0&=\Phi_1^\dagger\Phi_1+\Phi_2^\dagger\Phi_2,&\quad
K_1&=\Phi_1^\dagger\Phi_2+\Phi_2^\dagger\Phi_1,\\
K_2&=i\Phi_2^\dagger\Phi_1-i\Phi_1^\dagger\Phi_2,&\quad
K_3&=\Phi_1^\dagger\Phi_1-\Phi_2^\dagger\Phi_2,
\end{alignat}
and the four-vector
\bea
\tilde{\vec{K}}=(K_0,K_1,K_2,K_3)^\text{T}.
\eea
Then one can express the potential of the 2HDM as
\bea
V=\tilde{\vec{K}}^\text{T}\tilde{\vec{\xi}}+\tilde{\vec{K}}^\text{T}\tilde{E}\tilde{\vec{K}},
\eea
where
\bea
\tilde{\vec{\xi}}=
\begin{pmatrix}
	\xi_0 \\
	\vec{\xi}
\end{pmatrix},\quad
\tilde{E}=
\begin{pmatrix}
	\eta_{00} & \vec{\eta}^\text{T} \\
	\vec{\eta} & E
\end{pmatrix},
\eea
and in our notation\footnote{In \cite{Ferreira:2010yh} some of the potential parameters are defined slightly differently than ours.}
\begin{alignat}{2}
\xi_0&=-\frac{1}{4}(m_{11}^2+m_{22}^2), &\quad
\eta_{00}&=\frac{1}{8}(\lambda_1+\lambda_2)+\frac{1}{4}\lambda_3,\\
\vec{\xi}&=
\frac{1}{4}\begin{pmatrix}
	-2\,\Re m_{12}^2 \\
	2\,\Im m_{12}^2\\
	m_{22}^2-m_{11}^2
\end{pmatrix}, &\quad
\vec{\eta}&=
\frac{1}{4}\begin{pmatrix}
	\Re(\lambda_6+\lambda_7) \\
	-\Im(\lambda_6+\lambda_7)\\
	\half(\lambda_1-\lambda_2)
\end{pmatrix},
\end{alignat}
with
\begin{equation}
E=1/4\begin{pmatrix}
	\lambda_4+\Re\lambda_5 & -\Im\lambda_5 & \Re(\lambda_6-\lambda_7)\\
	-\Im\lambda_5 & \lambda_4-\Re\lambda_5 & -\Im(\lambda_6-\lambda_7)\\
	\Re(\lambda_6-\lambda_7) & -\Im(\lambda_6-\lambda_7) & \half(\lambda_1+\lambda_2)-\lambda_3
\end{pmatrix}.
\end{equation}
The authors of the paper \cite{Ferreira:2010yh} classified in their Table II
all\footnote{Only those symmetries of the scalar potential which could be extended to the whole Lagrangian of the model are discussed, so e.g., custodial symmetry has not been considered there as it would require no hypercharge coupling, $g'=0$. The custodial symmetry has been studied using the bilinear formalism
in~\cite{Grzadkowski:2010dj}.}  possible symmetries of the $\text{SU(2)}\times \text{U(1)}$ 2HDM potential in terms of the two vectors, $\vec{\xi}$ and $\vec{\eta}$, together with eigenvectors and eigenvalues of the three-by-three matrix $E$.
\subsection{Translating to the physical parameter set}
\label{Sec:Translating}
Our goal is to express the conditions for the different symmetries of the 2HDM in terms of constraints among the masses and couplings of the physical parameter set ${\cal P}$. For this purpose it is convenient to introduce the following vectors\footnote{Note that $\vec{F}_i^a$, $\vec{F}_i^b$ and $\vec{F}^c$ constitute a basis for $R^3$ in the case where $f_i\neq0$.}
\begin{subequations}
\bea
\vec{F}_i^a&\equiv&\left(f_i+f_i^*,i(f_i-f_i^*),0\right),\\
\vec{F}_i^b&\equiv&\left(-i(f_i-f_i^*),f_i+f_i^*,0\right),\\
\vec{F}^c&\equiv&\left(0,0,v\right).
\eea
\end{subequations}
Using the results from Appendix \ref{Translation} we find that in the Higgs basis
\bea
\vec{\xi}&=&-\frac{1}{4v^2}\left(
	e_1 M_1^2\vec{F}_1^a+e_2 M_2^2\vec{F}_2^a+e_3 M_3^2\vec{F}_3^a\right)\nonumber\\
	&&-\frac{1}{4v^3}
	\left[e_1^2 M_1^2+e_2^2 M_2^2+e_3^2 M_3^2+v^2(2 M_{H^\pm}^2-e_1 q_1-e_2 q_2-e_3 q_3)\right]\vec{F}^c
,\label{xi}\\
\vec{\eta}&=&\frac{1}{8v^4}\left[
	(e_1 M_1^2+v^2q_1)\vec{F}_1^a+(e_2 M_2^2+v^2q_2)\vec{F}_2^a+(e_3 M_3^2+v^2q_3)\vec{F}_3^a\right]\nonumber\\
	&&+\frac{1}{8v^5}
	(e_1^2 M_1^2+e_2^2 M_2^2+e_3^2 M_3^2-2 q v^4)\vec{F}^c.
\label{eta}
\eea
Likewise, one can translate the elements of the matrix $E$, yielding the following results, valid for the Higgs basis
\begin{subequations}
\bea
E_{11}&=&\frac{1}{8v^4}
\left[
(f_1+f_1^*)^2M_1^2
+(f_2+f_2^*)^2M_2^2+(f_3+f_3^*)^2M_3^2
-4v^2M_{H^\pm}^2
\right],\\
E_{12}&=&E_{21}=\frac{i}{8v^4}
\left[
(f_1^2-(f_1^*)^2)M_1^2
+(f_2^2-(f_2^*)^2)M_2^2
+(f_3^2-(f_3^*)^2)M_3^2
\right],\\
E_{13}&=&E_{31}=\frac{1}{8v^4}
\left[
(f_1+f_1^*)(e_1 M_1^2-v^2q_1)
+(f_2+f_2^*)(e_2M_2^2-v^2q_2)\right.\nonumber\\
&&\left.\hspace*{2cm}+(f_3+f_3^*)(e_3M_3^2-v^2q_3)
\right],\\
E_{22}&=&-\frac{1}{8v^4}
\left[
(f_1-f_1^*)^2M_1^2
+(f_2-f_2^*)^2M_2^2+(f_3-f_3^*)^2M_3^2
+4v^2M_{H^\pm}^2
\right],\\
E_{23}&=&E_{32}=\frac{i}{8v^4}
\left[
(f_1-f_1^*)(e_1M_1^2-q_1v^2)
+(f_2-f_2^*)(e_2M_2^2-q_2v^2)\right.\nonumber\\
&&\hspace*{2cm}\left.
+(f_3-f_3^*)(e_3M_3^2-q_3v^2)
\right],\\
E_{33}&=&\frac{1}{8v^4}
\left[
e_1^2M_1^2
+e_2^2M_2^2
+e_3^2M_3^2
\right]
+\frac{q}{4}-\frac{e_1q_1+e_2q_2+e_3q_3}{4v^2}.
\eea
\end{subequations}
Thus we see that by working in the Higgs basis, we managed to express $\vec{\xi}$, $\vec{\eta}$
 and $E$ in terms of the 11 parameters of $\pcal$ as well as the auxiliary quantities $f_i$ and $f_i^*$.
 We aim to express conditions for the different symmetries solely in terms of $\pcal$ (without $f_i$ and $f_i^*$). In Table II of \cite{Ferreira:2010yh}, the symmetry conditions are expressed in terms of the vanishing of one or more of the vectors $\vec\xi$, $\vec\eta$, $\vec{\xi}\times\vec{\eta}$, as well as how they align to the eigenvectors of the matrix $E$, and also the multiplicity of the eigenspaces. In order to analyze the vanishing of one of these vectors it is easier to study the vanishing of the squared norm of the vector, which turns out to be expressible in terms of $\pcal$ only (see next subsection). It also turns out that even if the vectors $\vec{\xi}$, $\vec{\eta}$ and $\vec{\xi}\times\vec{\eta}$, as well as the eigenvectors of $E$ are all dependent on $f_i$ and $f_i^*$, the conditions for obtaining a given symmetry, which depends on their relative alignment (parallel or perpendicular to each other) does not depend on $f_i$, and $f_i^*$. Thus, we are able to express the symmetry conditions in terms of $\pcal$ only, as expected.
\subsection{Properties of the vectors $\vec{\xi}$ and $\vec{\eta}$}
\label{Sec:Properties_vectors}
Both the vectors $\vec{\xi}$ and $\vec{\eta}$, as well as their cross product $\vec{\xi}\times\vec{\eta}$, will be needed in the discussion to follow. Also, we will need to formulate conditions for the vanishing of either of these vectors. For that purpose it is convenient to write out expressions for the squared length of each vector. Using (\ref{xi})--(\ref{eta}) along with (\ref{fifj}), first we find that
\bea
\vec{\xi}\times\vec{\eta}=\frac{1}{8v^4}
\left(
\chi_1\vec{F}_1^b
+\chi_2\vec{F}_2^b+\chi_3\vec{F}_3^b
-v^5\Im J_1\vec{F}^c
\right),
\eea
where $\chi_i=\left(\xi _3+2v^2 \eta _3\right)e_i M_i^2 +v^2\xi _3 q_i$, and $\Im J_1$ is a quantity encountered in the study of the CP properties of the 2HDM \cite{Lavoura:1994fv,Botella:1994cs,Gunion:2005ja,Grzadkowski:2014ada,Grzadkowski:2016szj}.
This quantity is part of a set of three invariant quantities, $\{J_1, J_2, J_3\}$, such that if all $\Im J_i = 0$ the 2HDM
vacuum preserves CP.\footnote{In fact, $J_3$ of the early papers \cite{Lavoura:1994fv,Botella:1994cs} corresponds to the present $\Im J_1$.}
$\Im J_1$ can be expressed in terms of the physical parameter set as
\bea
\Im J_1&=&\frac{1}{v^5}\sum_{i,j,k}\epsilon_{ijk}M_i^2e_ie_kq_j\nonumber\\
&=&\frac{1}{v^5}[M_1^2e_1(e_3q_2-e_2q_3)+M_2^2e_2(e_1q_3-e_3q_1)+M_3^2e_3(e_2q_1-e_1q_2)].
\eea
For the squared lengths of the vectors we then find
\bea
|\vec{\xi}|^2&=&\frac{1}{4v^4}\left[
e_1^2e_2^2(M_1^2-M_2^2)^2+e_1^2e_3^2(M_1^2-M_3^2)^2+e_2^2e_3^2(M_2^2-M_3^2)^2
\right]\nonumber\\
&&+\frac{1}{4}\left(
M_{H^\pm}^2-\frac{e_1q_1+e_2q_2+e_3q_3}{2}+\frac{e_1^2M_1^2+e_2^2M_2^2+e_3^2M_3^2}{2v^2}
\right)^2,\label{xisq}\\
|\vec{\eta}|^2&=&\frac{1}{16v^4}\left[
\left(e_1q_2-e_2q_1+\frac{e_1e_2}{v^2}(M_2^2-M_1^2)\right)^2+\left(e_1q_3-e_3q_1+\frac{e_1e_3}{v^2}(M_3^2-M_1^2)\right)^2\right.\nonumber\\
&&\left. \hspace*{1.3cm}
+\left(e_2q_3-e_3q_2+\frac{e_2e_3}{v^2}(M_3^2-M_2^2)\right)^2
\right]  \nonumber\\
&&+\frac{1}{16}\left(
q-\frac{e_1^2M_1^2+e_2^2M_2^2+e_3^2M_3^2}{2v^4}
\right)^2,\label{etasq}\\
|\vec{\xi}\times\vec{\eta}|^2&=&\frac{1}{16v^8}\left[
\left(e_1 e_2 \left(M_1^2-M_2^2\right) \left(\xi _3+2 \eta _3v^2\right)+\xi _3 v^2 \left(e_2 q_1-e_1 q_2\right)\right)^2\right.\nonumber\\
&&\hspace*{1cm}+\left(e_1 e_3 \left(M_1^2-M_3^2\right) \left(\xi _3+2 \eta _3 v^2\right)+\xi _3 v^2 \left(e_3 q_1-e_1 q_3\right)\right)^2\nonumber\\
&&\left.\hspace*{1cm}+\left(e_2 e_3 \left(M_2^2-M_3^2\right) \left(\xi _3+2 \eta _3 v^2\right)+\xi _3 v^2 \left(e_3 q_2-e_2 q_3\right)\right)^2
\right]\nonumber\\
&&+\frac{v^4}{64}\left(\Im J_1\right)^2.\label{cross-squared}
\eea
Let us at this point also introduce a shorthand notation for a quantity which we will encounter later in our study,
\bea
{\cal Q}^2\equiv (e_1q_2-e_2q_1)^2+(e_1q_3-e_3q_1)^2+(e_2q_3-e_3q_2)^2.
\eea
This quantity is always non-negative, and vanishes iff
\bea
(e_1q_2-e_2q_1)=(e_1q_3-e_3q_1)=(e_2q_3-e_3q_2)=0.
\eea

\subsection{The eigenvalues and eigenvectors of the matrix \texorpdfstring{$E$}{E}}
\label{Sec:Eigensystems}
In \cite{Ferreira:2010yh}, many of the symmetries we are about to discuss are formulated in terms of properties of the eigenvalues $\Lambda_m$ of the three-by-three matrix $E$, and the corresponding eigenvectors, $\vec{e}_m$:
\begin{equation}
E\,\vec{e}_m=\Lambda_m\,\vec{e}_m, \quad m=1,2,3.
\end{equation}
Note that $m\in\{1,2,3\}$ labels the eigenvalues, it should not be confused with the set $\{i,j,k\}$ used to label the three neutral scalars. Furthermore, the eigenvectors $\vec{e}_m$ should not be confused with the couplings $e_i$, $e_j$ or $e_k$. The characteristic equation of the matrix $E$ will be a cubic one, and in general we must express the roots of the characteristic equation using cube roots. In many of the physical configurations encountered, the characteristic equation factorizes and can be solved without the need for cube roots.
The discussion of the eigenvalues and the eigenvectors  of $E$ is relegated to Appendix \ref{Sec:Eigensystems_Appendix}.

\subsection{The vanishing of \texorpdfstring{$\Im J_1$}{Im J1}}
\label{Sec:Vanishing_imJ1}
Most symmetries we are about to discuss require that the cross product $\vec{\xi}\times\vec{\eta}$ vanishes. From (\ref{cross-squared}) we see that this will require $\Im J_1=0$. There are several ways for $\Im J_1$ to vanish, we need to explore them all. We list six physical configurations which together cover all situations under which $\Im J_1$ vanishes:
\begin{alignat}{2}
&\!\!\bullet &\ \
&\text{\bf Configuration 1: } \Im J_1= 0 \text{ because } M_1=M_2=M_3. \nonumber \\
&\!\!\bullet &\ \
&\text{\bf Configuration 2: } \Im J_1= 0 \text{ because } M_i=M_j,\, e_jq_i-e_iq_j=0,\nonumber\\
& & &\phantom{\text{\bf Configuration 2: }} \text{ and not Configuration 1.} \nonumber \\
&\!\!\bullet &\ \
&\text{\bf Configuration 3: } \Im J_1= 0 \text{ because } e_k=q_k=0, \text{ and not Configurations 1 or 2.} \nonumber \\
&\!\!\bullet &\ \
&\text{\bf Configuration 4: } \Im J_1= 0 \text{ because } M_i=M_j,\, e_k=0,\nonumber\\
& & &\phantom{\text{\bf Configuration 4: }} \text{ and not Configurations 1, 2 or 3.} \nonumber \\
&\!\!\bullet &\ \
&\text{\bf Configuration 5: } \Im J_1= 0 \text{ because } e_j=e_k=0, \text{ and not Configurations 1, 2, 3 or 4.} \nonumber \\
&\!\!\bullet &\ \
&\text{\bf Configuration 6: } \Im J_1= 0 \text{ because } q_k=\frac{e_je_kq_i(M_k^2-M_j^2)+e_ie_kq_j(M_i^2-M_k^2)}{e_ie_j(M_i^2-M_j^2)},\nonumber\\
& & &\phantom{\text{\bf Configuration 6: }} \text{ and not Configurations 1, 2, 3, 4 or 5.} \nonumber
\end{alignat}
We know from  earlier work \cite{Grzadkowski:2014ada} that Configurations 1--3 imply CP conservation for both the potential and the vacuum, since then all $\Im J_i=0$. Configurations 4--6 all imply $\Im J_1=0$, but some other $\Im J_i$ will be non-zero, thus CP is not conserved. The potential may still be CP invariant, in which case we will have spontaneous CP violation.
In  Configuration 6 it is implicitly understood that there is no mass degeneracy, and that all three gauge couplings $e_1$, $e_2$ and $e_3$ are non-vanishing.

\section{Results}
\label{sect:Results}
\setcounter{equation}{0}

In Table II of \cite{Ferreira:2010yh}, the six symmetry classes of the 2HDM, and the corresponding constraints
on the scalar potential parameters, are listed. At this point, we shall make note of the fact, that except for
a single constraint for the CP1 symmetry that requires $\vec{\xi}\times\vec{\eta}$ to be an eigenvector of $E$ (thus requiring $\vec{\xi}\times\vec{\eta}\neq\vec{0}$), all other constraints require $\vec{\xi}\times\vec{\eta}=\vec{0}$.
Therefore we may split the analysis into two parts---first we analyze how to get CP1 conservation when
$\vec{\xi}\times\vec{\eta}\neq\vec{0}$. Next, when continuing the analysis for the situations where
$\vec{\xi}\times\vec{\eta}=\vec{0}$ (both the second and the third option for CP1, as well as all the other
symmetries), we employ the fact that this also implies
$\Im J_1=0$, working systematically through the six configurations listed in section~\ref{Sec:Vanishing_imJ1}. The amount of configurations, sub-configurations and special configurations needed to be explored in order to arrive at the final results is substantial. We omit details of these calculations, however we believe that we have provided the reader with enough tools to explore and reproduce results listed hereafter on one's own, given the preliminary results in section~\ref{Sec:Preliminaries} and Appendix \ref{Sec:Eigensystems_Appendix}.

Hereafter we limit ourselves, unless explicitly stated, to cases with non-zero masses of non-Goldstone bosons.
\subsection{CP1 symmetry,}
\label{Sec:CP1}
If there exists a basis in which the potential of the 2HDM is invariant under the transformation
\bea
\Phi_i\rightarrow X_{ij} \Phi_j^*,
\eea
where $X=I_2$, then we say that the potential is invariant under CP1, i.e. there exists a basis in which the potential is invariant under complex conjugation, often referred to as ``standard" CP symmetry.
From Table II of \cite{Ferreira:2010yh} we see that the potential possesses the CP1 symmetry iff either of the following two conditions is met:
\begin{alignat}{2}
&\!\!\bullet &\ \
&\vec{\xi}\times\vec{\eta} \text{ is an eigenvector $\vec{e}$ of } E,\\
&\!\!\bullet &\ \
&\vec{\xi}\times\vec{\eta}=\vec{0};\quad\text{and}\quad \vec{\xi}\cdot\vec{{e}}=\vec{\eta}\cdot\vec{{e}}=0\text{ for some eigenvector } \vec{{e}} \text{ of } E.
\end{alignat}
Performing the analysis, we recover four already known \cite{Grzadkowski:2014ada, Grzadkowski:2016szj} cases when the 2HDM potential is CP conserving,
\begin{alignat}{2}
&\text{\bf Case \textoverline{A}:} &\ \
&  M_1=M_2=M_3.  \nonumber \\
&\text{\bf Case \textoverline{B}:} &\ \
&  M_i=M_j,\quad (e_jq_i-e_iq_j)=0.  \nonumber \\
&\text{\bf Case C:} &\ \
&  e_k=q_k=0. \nonumber\\
&\text{\bf Case D:} &\ \
& 2D M_{H^\pm}^2=v^2
\left[
e_1q_1M_2^2M_3^2+e_2q_2M_1^2M_3^2+e_3q_3M_1^2M_2^2-M_1^2M_2^2M_3^2
\right], \nonumber\\
& & & \phantom{\!\!\bullet}
2D q=
(e_2q_3-e_3q_2)^2M_1^2+(e_3q_1-e_1q_3)^2M_2^2+(e_1q_2-e_2q_1)^2M_3^2+M_1^2M_2^2M_3^2,
\nonumber
\end{alignat}
where the auxiliary sum $D$ is given by
\begin{equation}
D=e_1^2M_2^2M_3^2+e_2^2M_3^2M_1^2+e_3^2 M_1^2 M_2^2.
\end{equation}
Cases \textoverline{A}, \textoverline{B} and C are identical to what we have referred to as Configurations 1, 2 and 3 in section~\ref{Sec:Vanishing_imJ1}. From earlier work we know that these are cases under which we not only have a CP-invariant potential, but also a CP-invariant vacuum \cite{Grzadkowski:2014ada}.
Also from earlier work we know that the constraints of Case D only guarantee a CP-invariant potential, but the vacuum may or may not be CP-invariant, opening the possibility for having spontaneous CP violation \cite{Grzadkowski:2016szj}.

The reason for putting a bar over \textoverline{A} and \textoverline{B} is because these two cases of CP1 symmetry are unstable under the renormalization group equations. We shall in fact always put a bar over RGE unstable cases encountered, whereas the cases encountered that are RGE stable will be written without the bar. We have devoted section \ref{sec:rge} to a discussion of stability under RGE.

It is worth commenting on Case~\textoverline{B}, where mass degeneracy of the two fields $H_i$ and $H_j$ is accompanied by the constraint $(e_jq_i-e_iq_j)=0$ on the couplings. The mass degeneracy allows for an arbitrary angle\footnote{No physical observable may depend on this arbitrary angle. Thus, any quantity that depends on the arbitrary angle is unphysical, e.g. couplings which depend on this angle. The physical quantities will be the same regardless of which value we choose. Choosing to perform the calculations using an arbitrary angle will help us identifying which quantities are physical and which quantities are unphysical. The similarity to performing calculations in a particular gauge or in a general gauge is apparent.} in the neutral-sector rotation matrix allowing one to construct linear orthogonal combinations $H_a$ and $H_b$ out of $H_i$ and $H_j$.
In general, by a suitable rotation, one can arrange to make one of the new $H$'s even and the other odd under CP.

\begin{figure}[H]
	\centering
\includegraphics[height=3cm,angle=0]{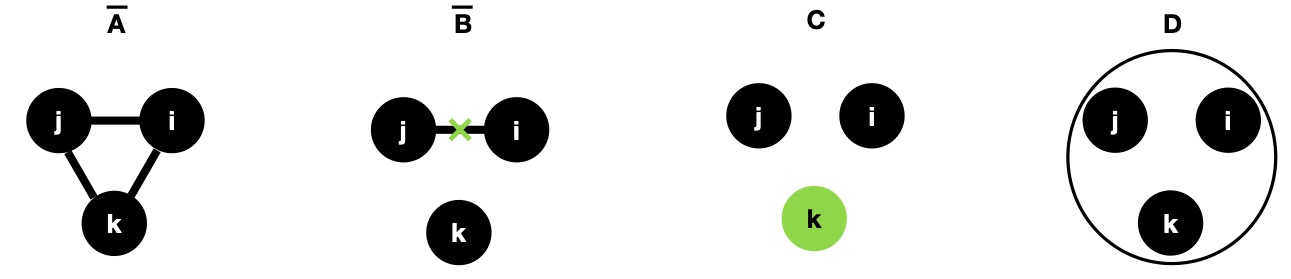}
	\caption{Visualization of the four cases of CP1. See text for explanation of the symbols.}
	\label{CP1figure}
\end{figure}

In Figure \ref{CP1figure} we make an attempt at visualizing the constraints of each of the four cases of CP1.
Each circle with a letter $i,j,k$ represents one of the three neutral scalars.
This graphic
representation is quite useful for a quick analysis of the several configurations of masses/couplings
yielding a given symmetry. Below we list some details of this convention:
\begin{itemize}
\item If a given circle is filled with a color (black or green), the corresponding scalar can have a non-vanishing mass. Scalars rendered massless by a symmetry will be represented as empty circles (see figure~\ref{U1figure} below).
\item Whenever there is a line connecting two circles, it means that the two connected neutral scalars are mass degenerate. {\it We then see in the visualization that Case} \textoverline{A} {\it corresponds to full mass degeneracy and Case} \textoverline{B} {\it to partial mass degeneracy}.
\item Whenever there is a green cross on a line connecting two mass degenerate neutral scalars labeled $i$ and $j$, this tells us that, in addition to the mass degeneracy, the constraint $e_jq_i-e_iq_j=0$ applies to the couplings of those two scalars. {\it This is seen in the visualization of Case} \textoverline{B}.
\item Whenever a circle is filled with green color (rather than black) this means that the corresponding neutral scalar does
not couple to $ZZ$, $W^+W^-$ or $H^+H^-$ pairs. {\it This illustrates, for Case C, that $e_k=q_k=0$}.
\item Whenever the circles representing the three neutral scalars are enclosed by a larger circle ({\it as shown for Case D}), this means that the two constraints characterizing Case D apply (one constraint on $\mchsq$ and another on $q$).
\end{itemize}

As for the remaining symmetries, $Z_2$, U(1), CP2, CP3 and SO(3), we know that if the potential respects any
one of them, it will also be CP1 symmetric.
The vacuum will be CP-invariant as well in any of
those cases\footnote{Remember that we are only dealing with {\it exact} symmetries and are not considering
soft breaking terms in the potential.}, indicating that Configurations 1, 2 and 3 are the only
configurations that need to be explored for the remaining symmetries. This also means that they will all have
to satisfy at least the constraints of cases \textoverline{A}, \textoverline{B} or C.
\subsection{\texorpdfstring{$Z_2$}{Z2} symmetry}
\label{Sec:Z_2}
If there exists a basis in which the potential of the 2HDM is invariant under the transformation
\bea
\Phi_1\rightarrow \Phi_1,\, \Phi_2\rightarrow -\Phi_2,
\eea
then we say that the potential is invariant under $Z_2$.
From Table II of \cite{Ferreira:2010yh} we see that the potential possesses the $Z_2$ symmetry iff the following condition is met
\begin{alignat}{2}
&\!\!\bullet &\ \
&\vec{\xi}\times\vec{e}=\vec{\eta}\times\vec{e}=\vec{0}\text{ for some eigenvector } \vec{{e}} \text{ of } E.
\end{alignat}
We find a total of six cases when the 2HDM potential is $Z_2$ invariant\footnote{These six cases will also follow from using the commutators presented in \cite{Davidson:2005cw}, guaranteeing a $Z_2$-symmetric potential if they all vanish.},
\begin{alignat}{2}
&\text{\bf Case \textoverline{AD}:} &\ \
&M_1=M_2=M_3,\nonumber\\
& & & 2M_{H^\pm}^2=e_1q_1+e_2q_2+e_3q_3-M_1^2,\quad 2M_1^2v^2q={\cal Q}^2+M_1^4.\nonumber\\
&\text{\bf Case \textoverline{ABBB}:} &\ \
&M_1=M_2=M_3,\quad
{\cal Q}^2=0.\nonumber\\
&\text{\bf Case \textoverline{BD}:} &\ \
&M_i=M_j,\quad (e_iq_j-e_jq_i)=0,\nonumber\\
& & & \phantom{\!\!\bullet}
2\left[e_k^2M_i^2+(e_i^2+e_j^2)M_k^2\right]M_{H^\pm}^2=v^2\left[(e_iq_i+e_jq_j-M_i^2)M_k^2+e_kq_kM_i^2\right],\nonumber\\
& & & \phantom{\!\!\bullet} 2\left[e_k^2M_i^2+(e_i^2+e_j^2)M_k^2\right]q=(e_iq_k-e_kq_i)^2+(e_jq_k-e_kq_j)^2+M_i^2M_k^2.\nonumber\\
&\text{\bf Case \textoverline{BC}:} &\ \
&M_i=M_j,\quad (e_iq_j-e_jq_i)=0,\quad
e_k=q_k=0.\nonumber\\
&\text{\bf Case CD:} &\ \
&e_k=q_k=0, \quad 2(e_j^2M_i^2+e_i^2M_j^2)M_{H^\pm}^2=v^2(e_jq_jM_i^2+e_iq_iM_j^2-M_i^2M_j^2),\nonumber\\
& & & \phantom{\!\!\bullet}
2(e_j^2M_i^2+e_i^2M_j^2)q=(e_jq_i-e_iq_j)^2+M_i^2M_j^2.\nonumber\\
&\text{\bf Case CC:} &\ \
&e_j=q_j=e_k=q_k=0.\nonumber
\end{alignat}

\begin{figure}[H]
	\centering
\includegraphics[height=3cm,angle=0]{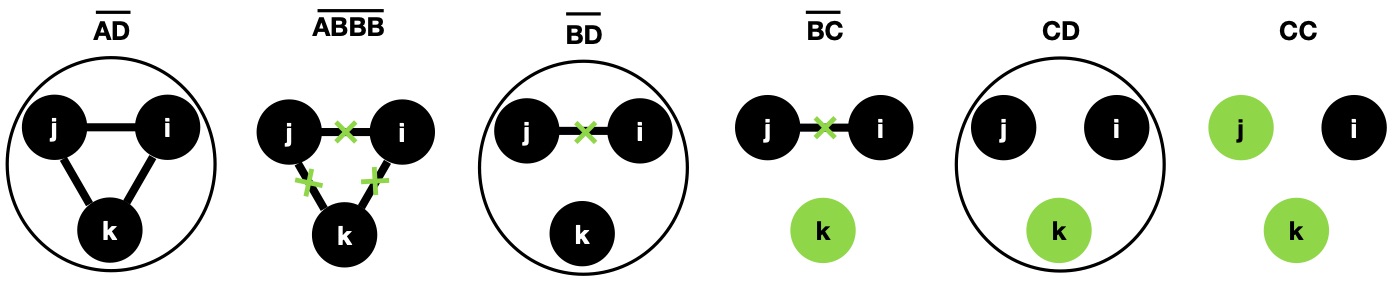}
	\caption{Visualization of the cases of $Z_2$.}
	\label{Z2figure}
\end{figure}

In Figure \ref{Z2figure} we visualize the constraints for each of the six cases of $Z_2$. The namings of these six cases are related to the four individual cases of CP1. We  note that each of the six cases of $Z_2$ is obtained by simultaneously  imposing two or more of the conditions yielding CP1 (this is in agreement with theorem 1 of \cite{Ferreira:2010yh}):
\begin{trivlist}
\item
$\!\!\bullet$ Case \textoverline{AD} is the combination of cases \textoverline{A} and D.
\item
$\!\!\bullet$ Case \textoverline{ABBB} is the combination of three different cases \textoverline{B} (with different pairs of indices), also satisfying case \textoverline{A}.
\item
$\!\!\bullet$ Case \textoverline{BD} is the combination of cases \textoverline{B} and D.
\item
$\!\!\bullet$ Case \textoverline{BC} is the combination of cases \textoverline{B} and C.
\item
$\!\!\bullet$ Case CD is the combination of cases C and D.
\item
$\!\!\bullet$ Case CC is the combination of two different cases C (with different indices).
\end{trivlist}
From this labeling it is apparent that all cases of $Z_2$ satisfy at least the constraints of cases \textoverline{A}, \textoverline{B} or C as stated at the end of the previous subsection.

It follows from the discussion of Case~\textoverline{B} above that in the fully degenerate Case \textoverline{ABBB} two linear combinations (call them $H_a$ and $H_b$) could be formed out of $H_i$,  $H_j$ and $H_k$ that both decouple from  gauge bosons ($e_a=e_b=0$) and from the charged scalars ($q_a=q_b=0$), while the third one carries the full-strength couplings. The two states that decouple have opposite CP.
\subsection{U(1) symmetry}
\label{Sec:U(1)}
If there exists a basis in which the potential of the 2HDM is invariant under the transformation
\bea
\Phi_1\rightarrow e^{-i\theta}\Phi_1,\, \Phi_2\rightarrow e^{i\theta}\Phi_2
\label{u1def}
\eea
for an arbitrary angle $\theta$, then we say that the potential is invariant under U(1).
From Table~II of \cite{Ferreira:2010yh} we see that the potential possesses the U(1) symmetry iff either of the following two conditions is met:
\begin{alignat}{2}
&\!\!\bullet &\ \
&\Delta=0;\quad\text{and}\quad \vec{\xi}\times\vec{e}=\vec{\eta}\times\vec{e}=\vec{0} \text{ where $\vec{e}$ is an eigenvector from a one-dimensional}\nonumber\\
&\phantom{\!\!\bullet }&\ \
&\text{eigenspace of } E, \\
&\!\!\bullet &\ \
&\Delta=0;\, \Delta_0=0;\quad\text{and}\quad  \vec{\xi}\times\vec{\eta}=\vec{0},
\end{alignat}
where $\Delta$ and $\Delta_0$ are defined in appendix \ref{Sec:Eigensystems_Appendix}.
The condition $\Delta = 0$ implies that the $E$ matrix has two degenerate eigenvalues,
and requiring $\Delta_0 = 0$ as well, $E$ will have three degenerate eigenvalues.

We find a total of four cases when the 2HDM potential is U(1) invariant,
\begin{alignat}{2}
&\text{\bf Case \textoverline{ABBB}:} &\ \
&M_1=M_2=M_3,\quad
{\cal Q}^2=0.\nonumber\\
&\text{\bf Case \textoverline{B$_0$D}:} &\ \
&M_i=M_j=0,\quad (e_iq_j-e_jq_i)=0,\quad
2(e_i^2+e_j^2)M_{H^\pm}^2=v^2(e_iq_i+e_jq_j),\nonumber\\
& & & \phantom{\!\!\bullet} 2(e_i^2+e_j^2)M_k^2q=(e_iq_k-e_kq_i)^2+(e_jq_k-e_kq_j)^2.\nonumber\\
&\text{\bf Case BCC:} &\ \
&M_j=M_k,\quad
e_j=q_j=e_k=q_k=0.\nonumber\\
&\text{\bf Case C$_0$D:} &\ \
&e_k=q_k=0, \quad 2(e_j^2M_i^2+e_i^2M_j^2)M_{H^\pm}^2=v^2(e_jq_jM_i^2+e_iq_iM_j^2-M_i^2M_j^2),\nonumber\\
& & & \phantom{\!\!\bullet}
2(e_j^2M_i^2+e_i^2M_j^2)q=(e_jq_i-e_iq_j)^2+M_i^2M_j^2,\quad M_k=0.\nonumber
\end{alignat}

\begin{figure}[H]
	\centering
\includegraphics[height=3cm,angle=0]{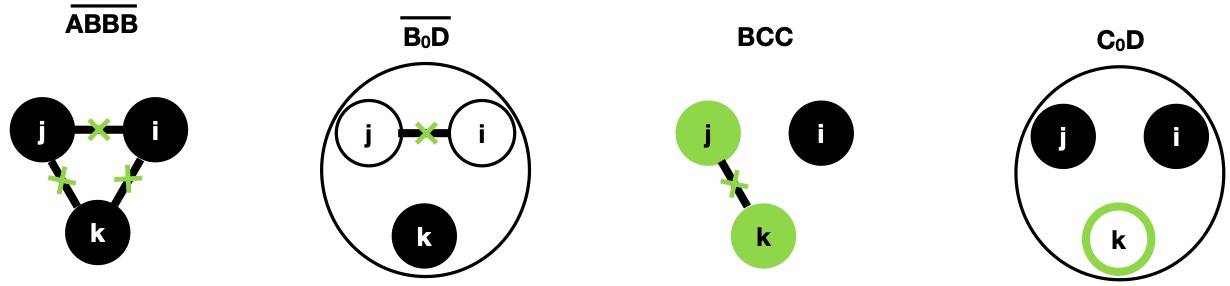}
	\caption{Visualization of the cases of U(1).}
	\label{U1figure}
\end{figure}

In Figure \ref{U1figure} we visualize the constraints for each of the four cases of U(1). We follow the same pattern as before when giving names to these cases, with the additional subscript "0" whenever there are neutral scalars with vanishing masses.
All U(1) invariant potentials are also $Z_2$ invariant,
as is seen when comparing Figures~\ref{Z2figure} and \ref{U1figure}.
\subsection{CP2 symmetry}
\label{Sec:CP2}
If there exists a basis in which the potential of the 2HDM is invariant under the transformation
\bea
\Phi_i\rightarrow X_{ij} \Phi_j^*,\quad \text{where}\quad
X=\begin{pmatrix}
0 & 1 \\
-1 & 0
\end{pmatrix},
\eea
then we say that the potential is invariant under CP2\,\footnote{The CP2 symmetry cannot be extended to the fermion sector in an acceptable manner, as it always implies
at least one massless family~\cite{Maniatis:2007de,Maniatis:2009vp,Ferreira:2010bm}.}.
From Table II of \cite{Ferreira:2010yh} we see that the potential possesses the CP2 symmetry iff the following condition is met:
\begin{alignat}{2}
&\!\!\bullet &\ \
&(\vec{\xi},\vec{\eta})=(\vec{0},\vec{0}).
\end{alignat}
We find a total of three cases when the 2HDM potential is CP2 invariant,
\begin{alignat}{2}
&\text{\bf Case \textoverline{ABBBD}:} &\ \
&M_1=M_2=M_3,\quad
{\cal Q}^2=0,\nonumber\\
& & &2M_{H^\pm}^2=e_1q_1+e_2q_2+e_3q_3-M_1^2,\quad 2v^2q=M_1^2.\nonumber\\
&\text{\bf Case \textoverline{BCD}:} &\ \
&M_i=M_j,\quad (e_iq_j-e_jq_i)=0,\nonumber\\
& & & 2M_{H^\pm}^2=e_iq_i+e_jq_j-M_i^2,\quad 2v^2q=M_i^2,\quad
e_k=q_k=0.\nonumber\\
&\text{\bf Case CCD:} &\ \
&e_j=q_j=e_k=q_k=0,\quad 2M_{H^\pm}^2=e_iq_i-M_i^2,\quad 2v^2q=M_i^2.\nonumber
\end{alignat}

\begin{figure}[H]
	\centering
\includegraphics[height=3cm,angle=0]{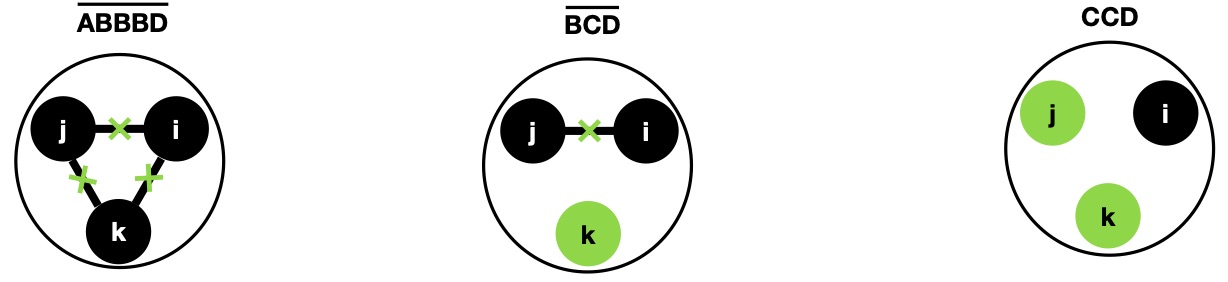}
	\caption{Visualization of the cases of CP2.}
	\label{CP2figure}
\end{figure}

In Figure \ref{CP2figure} we visualize the constraints for each of the three cases of CP2. All CP2 invariant potentials are also $Z_2$ invariant, as is seen when comparing Figures~\ref{Z2figure} and \ref{CP2figure}.

\subsection{CP3 symmetry}
\label{Sec:CP3}
If there exists a basis in which the potential of the 2HDM is invariant under the transformation
\bea
\Phi_i\rightarrow X_{ij} \Phi_j^*,\quad \text{where}\quad
X=\begin{pmatrix}
	\cos\theta & \sin\theta \\
	-\sin\theta & \cos\theta
\end{pmatrix},
\eea
for any $0<\theta<\pi/2$,
then we say that the potential is invariant under CP3\footnote{The only viable extension of CP3 to the fermion sector occurs for $\theta = \pi/3$, any other choice of angle implies a massless family~\cite{Ferreira:2010bm}.}.
From Table~II of \cite{Ferreira:2010yh} we see that the potential possesses the CP3 symmetry iff
the following condition is met
\begin{alignat}{2}
&\!\!\bullet &\ \
&\Delta=0;\quad\text{with}\quad(\vec{\xi},\vec{\eta})=(\vec{0},\vec{0}).
\end{alignat}
We find a total of five cases when the 2HDM potential is CP3 invariant,
\begin{alignat}{2}
&\text{\bf Case \textoverline{ABBBD}:} &\ \
&M_1=M_2=M_3,\quad
{\cal Q}^2=0,\nonumber\\
& & &2M_{H^\pm}^2=e_1q_1+e_2q_2+e_3q_3-M_1^2,\quad 2v^2q=M_1^2.\nonumber\\
&\text{\bf Case \textoverline{B$_0$CD}:} &\ \
&M_i=M_j=0,\quad (e_iq_j-e_jq_i)=0,\nonumber\\
& & & 2M_{H^\pm}^2=e_iq_i+e_jq_j,\quad q=0,\quad
e_k=q_k=0.\nonumber\\
&\text{\bf Case \textoverline{BC$_0$D}:} &\ \
&M_i=M_j,\quad (e_iq_j-e_jq_i)=0,\nonumber\\
& & & 2M_{H^\pm}^2=e_iq_i+e_jq_j-M_i^2,\quad 2v^2q=M_i^2,\quad
e_k=q_k=0,\quad M_k=0.\nonumber\\
&\text{\bf Case BCCD:} &\ \
&e_j=q_j=e_k=q_k=0,\quad 2M_{H^\pm}^2=e_iq_i-M_i^2,\quad 2v^2q=M_i^2,\quad M_j=M_k.\nonumber\\
&\text{\bf Case C$_0$CD:} &\ \
&e_j=q_j=e_k=q_k=0,\quad 2M_{H^\pm}^2=e_iq_i-M_i^2,\quad 2v^2q=M_i^2,\quad M_j=0.\nonumber
\end{alignat}

\begin{figure}[H]
	\centering
\includegraphics[height=3cm,angle=0]{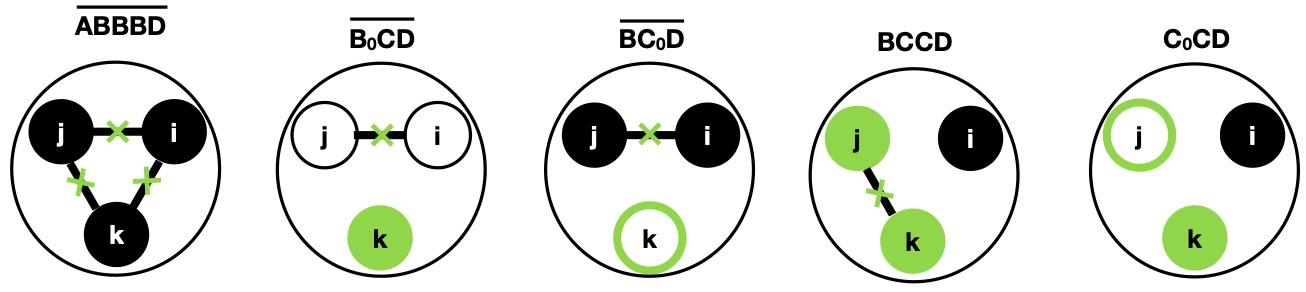}
	\caption{Visualization of the cases of CP3.}
	\label{CP3figure}
\end{figure}
In Figure \ref{CP3figure} we visualize the constraints for each of the five cases of CP3. All CP3 invariant potentials are also both U(1) invariant and CP2 invariant, as is seen when comparing Figures~\ref{U1figure}, \ref{CP2figure} and \ref{CP3figure}.

\subsection{SO(3) symmetry}
\label{Sec:SO(3)}
If there exists a basis in which the potential of the 2HDM is invariant under the transformation
\bea
\Phi_i\rightarrow U_{ij} \Phi_j,
\eea
where $U$ is any U(2) matrix, then we say that the potential is invariant under SO(3)\footnote{It might seem that
the symmetry group involved in these transformations would be the full U(2), but as argued
in \cite{Ivanov:2006yq,Ivanov:2007de}, taking into account the U(1) hypercharge symmetry underlying the
theory, the largest Higgs-family symmetry is indeed SO(3). This is particularly evident in the bilinear formalism.}.
From Table II of \cite{Ferreira:2010yh} we see that the potential possesses the SO(3) symmetry iff the following conditions are met:
\begin{alignat}{2}
&\!\!\bullet &\ \
&\Delta=0;\quad \Delta_0=0;\quad\text{with}\quad(\vec{\xi},\vec{\eta})=(\vec{0},\vec{0}).
\end{alignat}
We find a total of two cases when the 2HDM potential is SO(3) invariant,
\begin{alignat}{2}
&\text{\bf Case \textoverline{A$_0$B$_0$B$_0$B$_0$D}:} &\ \
&M_1=M_2=M_3=0,\quad
{\cal Q}^2=0,\nonumber\\
& & &2M_{H^\pm}^2=e_1q_1+e_2q_2+e_3q_3,\quad q=0.\nonumber\\
&\text{\bf Case B$_0$C$_0$C$_0$D:} &\
&M_j=M_k=0,\quad e_j=q_j=e_k=q_k=0,\nonumber\\
& & &2M_{H^\pm}^2=e_iq_i-M_i^2,\quad 2v^2q=M_i^2.\nonumber
\end{alignat}
\begin{figure}[H]
	\centering
\includegraphics[height=3cm,angle=0]{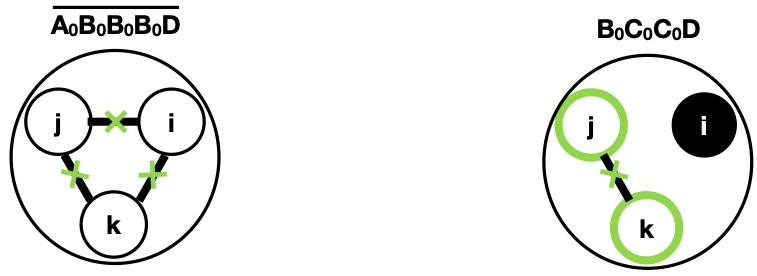}
	\caption{Visualization of the cases of SO(3).}
	\label{SO3figure}
\end{figure}
In Figure \ref{SO3figure} we visualize the constraints of each of the two cases of SO(3). All SO(3) invariant potentials are also CP3 invariant, as is seen when comparing Figures~\ref{CP3figure} and \ref{SO3figure}.

\section{Analysis}
\label{Sec:analysis}
\setcounter{equation}{0}
We shall now demonstrate explicitly how the different cases presented in the previous section can be realized for specific choices of the potential, together with a suitable basis. That is, we find the explicit contraints on the potential parameters and the vacuum parameters that correspond to each of the cases presented. Thus, we will see that each of the cases presented as contraints on masses and couplings is in fact realizable by explicit construction of potential and vacuum.

For each of the symmetry classes we write out the most general potential for which the symmetry is manifest, along with a vacuum written out for a general basis. Next, we write out the resulting stationary-point conditions, which often can be solved in more than one way. Some solutions imply one of the doublets having a VEV equal to zero, for others both
doublets have non-zero VEVs.  Different ways of solving the stationary-point equations often lead to the manifestation of different cases within each symmetry class. The cases in which the symmetry is manifest in a Higgs basis versus the cases in which the symmetry is manifest in a non-Higgs basis often (but not always) depends upon whether the symmetry under consideration is spontaneously broken or not.

We will also stress that these are tree-level classifications. As we will discuss in the next section, some of the constraints corresponding to the cases listed in the previous section are not stable under the RGE. However, since the symmetry under consideration is preserved also by the higher-order effects, we must necessarily remain within the same symmetry class. We return to this issue in section~\ref{sec:rge}.

\subsection{CP1 symmetry}
If the potential is invariant under CP1, there exists a basis in which all the parameters of the potential
are real. The VEV can either be real (CP conservation) or complex (spontaneous CP violation). Cases \textoverline{A},
\textoverline{B} and C correspond to
complete CP conservation (potential and vacuum) while case D allows for spontaneous CP violation.
These are known results \cite{Grzadkowski:2014ada,Grzadkowski:2016szj} so we do not repeat the details.
Anticipating results of section~\ref{sec:rge}, we will find that cases~\textoverline{A} and \textoverline{B}, involving mass degeneracies, are not RGE stable, these will ``migrate'' into case C which is RGE stable at the one-loop level.

Since hereafter we are going to discuss vacuum structure and the possibility of spontaneous symmetry breaking,  we switch from the Higgs basis to a generic one, as this is more convenient for the discussion.

\subsection{\texorpdfstring{$Z_2$}{Z2} symmetry}

If the potential is invariant under $Z_2$, there exists a basis in which $m_{12}^2=0$ and $\lambda_6=\lambda_7=0$.
Writing out the potential in such a basis, it is given by
\begin{align}
\label{Eq:potZ2}
V(\Phi_1,\Phi_2) &= -\frac12\left\{m_{11}^2\Phi_1^\dagger\Phi_1
+ m_{22}^2\Phi_2^\dagger\Phi_2 \right\} \nonumber \\
& + \frac{\lambda_1}{2}(\Phi_1^\dagger\Phi_1)^2
+ \frac{\lambda_2}{2}(\Phi_2^\dagger\Phi_2)^2
+ \lambda_3(\Phi_1^\dagger\Phi_1)(\Phi_2^\dagger\Phi_2)
+ \lambda_4(\Phi_1^\dagger\Phi_2)(\Phi_2^\dagger\Phi_1)\nonumber \\
&+ \frac12\left[\lambda_5(\Phi_1^\dagger\Phi_2)^2 + \hc\right].
\end{align}
Since we do not know the form of the vacuum, we shall assume the most general charge-conserving form, and parametrize the Higgs doublets as
\begin{equation}
\Phi_j=e^{i\xi_j}\left(
\begin{array}{c}\varphi_j^+\\ (v_j+\eta_j+i\chi_j)/\sqrt{2}
\end{array}\right), \quad
j=1,2.\label{vevsZ2}
\end{equation}
Here $v_j$ are real numbers, so that $v_1^2+v_2^2=v^2$. The fields $\eta_j$ and $\chi_j$ are real, whereas $\varphi_j^+$ are complex fields.
Then the most general form of the vacuum reads
\bea
\left<\Phi_j\right>=\frac{e^{i\xi_j}}{\sqrt{2}}\left(
\begin{array}{c}0\\ v_j
\end{array}\right).
\eea
Note that the phases $\xi_j$ are extracted from the whole doublet, not from the VEVs only.

Next, let us define orthogonal states
\beq
\left(
\begin{array}{c}G_0\\ \eta_3
\end{array}\right)
=
\left(
\begin{array}{cc}v_1/v & v_2/v\\ -v_2/v & v_1/v
\end{array}\right)
\left(
\begin{array}{c}\chi_1\\ \chi_2
\end{array}\right)
\eeq
and
\beq
\left(
\begin{array}{c}G^\pm\\ H^\pm
\end{array}\right)
=
\left(
\begin{array}{cc}v_1/v & v_2/v\\ -v_2/v & v_1/v
\end{array}\right)
\left(
\begin{array}{c}\varphi_1^\pm\\ \varphi_2^\pm
\label{orthogonal}
\end{array}\right),
\eeq
so that $G_0$ and $G^\pm$ become the massless Goldstone fields and $H^\pm$ are the charged scalars. The neutral fields $\eta_i$ are related to the mass-eigenstate fields $H_i$ by (\ref{Eq:R-def}). The masses of the physical scalars are read off from the bilinear terms in the potential and, using (\ref{Eq:cal-M}), the mass-squared matrix (which now has a different form than in the Higgs basis) is diagonalized.

Without loss of generality we can rephase $\Phi_2$ in order to make $\lambda_5$ real. Next, utilizing a simultaneous rephasing (by the weak hypercharge) of both $\Phi_1$ and $\Phi_2$ we can make $\left<\Phi_1\right>$ real and non-negative.

This leaves us with the following potential
\begin{align}
\label{Eq:potZ2rephased}
V(\Phi_1,\Phi_2) &= -\frac12\left\{m_{11}^2\Phi_1^\dagger\Phi_1
+ m_{22}^2\Phi_2^\dagger\Phi_2 \right\} \nonumber \\
& + \frac{\lambda_1}{2}(\Phi_1^\dagger\Phi_1)^2
+ \frac{\lambda_2}{2}(\Phi_2^\dagger\Phi_2)^2
+ \lambda_3(\Phi_1^\dagger\Phi_1)(\Phi_2^\dagger\Phi_2)
+ \lambda_4(\Phi_1^\dagger\Phi_2)(\Phi_2^\dagger\Phi_1)\nonumber \\
&+ \frac{\Re\lambda_5}{2}\left\{(\Phi_1^\dagger\Phi_2)^2+(\Phi_2^\dagger\Phi_1)^2\right\},
\end{align}
and the vacuum
\bea
\left<\Phi_1\right>=\frac{1}{\sqrt{2}}\left(
\begin{array}{c}0\\ v_1
\end{array}\right),\,
\left<\Phi_2\right>=\frac{e^{i\xi}}{\sqrt{2}}\left(
\begin{array}{c}0\\ v_2
\end{array}\right),\label{vacuum}
\eea
with $v_i\geq0$.

Minimizing the potential with respect to the fields yields the stationary-point equations
\bea \label{Eq:Z2-minimization}
v_1\left[-m_{11}^2+v_1^2\lambda_1+v_2^2(\lambda_3+\lambda_4+c_{2\xi}\Re\lambda_5)\right]&=&0,\nonumber\\
v_2\left[-m_{22}^2+v_2^2\lambda_2+v_1^2(\lambda_3+\lambda_4+c_{2\xi}\Re\lambda_5)\right]&=&0,\nonumber\\
v_1v_2s_{2\xi}\Re\lambda_5&=&0.
\eea
We shall assume that $\Re\lambda_5\neq0$, or otherwise we would have a U(1)-symmetric potential, which we shall treat later.
Allowing for solutions with one vanishing VEV, the stationary-point equations can be solved by simply putting
\bea
v_2=0,\quad
m_{11}^2=v^2\lambda_1.
\label{eq:m11}
\eea
This solution corresponds to a $Z_2$-symmetric vacuum. There is no need to also consider $v_1=0$, since this is related to $v_2=0$ simply by an interchange of the two doublets.

Another solution of the stationary-point equations can be found whenever both $v_i\neq0$. Then the stationary-point equations are solved by
\bea
m_{11}^2&=&v_1^2\lambda_1+v_2^2(\lambda_3+\lambda_4+c_{2\xi}\Re\lambda_5),\nonumber\\
m_{22}^2&=&v_2^2\lambda_2+v_1^2(\lambda_3+\lambda_4+c_{2\xi}\Re\lambda_5),\nonumber\\
s_{2\xi}&=&0.
\eea
This solution corresponds to a spontaneously broken $Z_2$-symmetry.
We see that these two ways of satisfying the stationary-point equations are topologically different, meaning one cannot get from the spontaneously broken vacuum solution to the $Z_2$-symmetric vacuum solution simply by letting $v_2\to0$ in a continuous way. These two situations will therefore lead to different physics, as we will now see.
\subsubsection{\texorpdfstring{$Z_2$}{Z2} symmetric potential and vacuum}
\label{sec:cc}
With the potential of (\ref{Eq:potZ2rephased}) and a vacuum of the form
\bea
\left<\Phi_1\right>=\frac{1}{\sqrt{2}}\left(
\begin{array}{c}0\\ v
\end{array}\right),\quad
\left<\Phi_2\right>=\left(
\begin{array}{c}0\\ 0
\end{array}\right),
\eea
we have obtained the potential and the vacuum of the Inert Doublet Model
(IDM)~\cite{Ma:1978,Barbieri:2006,Cao:2007,LopezHonorez:2006}. In this model the $Z_2$
symmetry is preserved by the vacuum, and the lightest neutral scalar from the
second doublet becomes a viable dark matter candidate.

With a vanishing VEV, the phase $\xi$ can be absorbed into $\Phi_2$ and the parametrization of the doublets becomes
\begin{equation}
\Phi_1=\left(
\begin{array}{c}G^+\\ (v+G_0+iH_1)/\sqrt{2}
\end{array}\right), \quad
\Phi_2=\left(
\begin{array}{c}H^+\\ (H_2+iH_3)/\sqrt{2}
\end{array}\right).\label{vevsZ2symm}
\end{equation}
This is equivalent to simply putting $\xi=0$.

The mass-squared matrix becomes
\bea
{\cal M}^2=v^2\text{diag}\left[\lambda_1,\half\left(\lambda_3+\lambda_4+\Re\lambda_5-\frac{m_{22}^2}{v^2}\right),\half\left(\lambda_3+\lambda_4-\Re\lambda_5-\frac{m_{22}^2}{v^2}\right)\right]\label{eq:mmat},
\eea
and the charged mass is given in (\ref{chargedmass}),
\bea
\mchsq=\frac{1}{2}\left(v^2\lambda_3-m_{22}^2\right).
\eea
\paragraph{No mass degeneracy (RGE stable)}\mbox{}\\
Provided there is no mass degeneracy, the rotation matrix for the neutral sector is simply
the three-by-three unit matrix, so that the relevant couplings become
\bea
e_1=v,\quad
e_2=e_3=0,\quad
q_1=v\lambda_3,\quad
q_2=q_3=0.\label{Eq:IDM-couplings}
\eea
This is then seen to be a realization of Case CC.
\paragraph{Partial mass degeneracy $M_1=M_2$ (RGE unstable)}\mbox{}\\
We will not yet allow for mass degeneracy between $M_2$ and $M_3$ or full mass degeneracy, since this would require $\lambda_5=0$ which yields a U(1) symmetric model (which will be treated later). Allowing for mass degeneracy\footnote{Mass degeneracy between $M_1$ and $M_3$ yields a similar result.} between $M_1$ and $M_2$  requires $m_{22}^2=(-2\lambda_1+\lambda_3+\lambda_4+\Re\lambda_5)v^2$.
The mass-squared matrix now becomes
\bea
{\cal M}^2=v^2\text{diag}(\lambda_1,\lambda_1,\lambda_1-\Re\lambda_5),
\eea
and the charged mass
\bea
\mchsq=\frac{1}{2}v^2\left(2\lambda_1-\lambda_4-\Re\lambda_5\right).
\eea
One may now argue that the neutral-sector mass matrix is already diagonal, and the rotation matrix will simply be the unit matrix. While this argument is notably correct, one can also (because of the mass degeneracy present) use the rotation matrix
\bea
R=\begin{pmatrix}
	c_\alpha    &  s_\alpha   & 0   \\
	-s_\alpha   &  c_\alpha  & 0   \\
	0    &  0   & 1
\end{pmatrix},
\label{eq:roa}
\eea
where $\alpha$ is a completely arbitrary angle to preserve the diagonal structure of the mass matrix, but the mass eigenstate fields will have a different admixture of the fields $H_1$ and $H_2$ for each value of $\alpha$, affecting the couplings involving the neutral scalars. In particular, using (\ref{eq:roa}), we find the following couplings
\bea
e_1=v\cos \alpha,\quad
e_2=-v\sin\alpha,\quad
q_1=v\cos\alpha\lambda_3,\quad
q_2=-v\sin\alpha\lambda_3,\quad
e_3=q_3=0,\label{BCcouplings}
\eea
some of which depend on the arbitrary angle $\alpha$.
One can easily see that for any value of $\alpha$, $e_1q_2-e_2q_1=0$. Thus, this is a realization of Case \textoverline{BC}.

It should be emphasized that the arbitrariness of the angle $\alpha$ has nothing to do with basis freedom, it is only an artifact resulting from the mass degeneracy. Physics cannot depend upon the value of $\alpha$, all physical observables must in this mass-degenerate case be independent of $\alpha$.
Since the couplings $e_1$, $e_2$, $q_1$ and $q_2$ all depend on $\alpha$, none of these couplings are physical. Neither can they be made physical simply by picking a particular value of the unphysical $\alpha$. However, combinations of these couplings like $e_1^2+e_2^2=v^2$ or $q_1^2+q_2^2=v^2\lambda_3^2$ or $e_1q_1+e_2q_2=v^2\lambda_3$ are independent of $\alpha$ and are physical. Thus, in processes with external $H_1$ and $H_2$ one should sum corresponding squares of amplitudes (no interference)\footnote{Consider $H^+H^-\to H_{1,2}$ (external). The sum of squared amplitudes becomes proportional to $q_1^2+q_2^2$, which is physical since it is independent of $\alpha$.}, while in the case of virtual $H_{1,2}$, summation should be made at the level of amplitudes\footnote{Consider $W^+W^-\to H_{1,2,3}\text{ (virtual) }\to H^+H^-$. The amplitude becomes proportional to $e_1q_1+e_2q_2$, which is physical since it is independent of $\alpha$.}. In the end, any dependence on the mixing angle $\alpha$ must vanish.
One possible construction is to define one field $H_a$ carrying the full-strength interaction, $e_a=v$ and $q_a=v\lambda_3$, and another field $H_b$, that decouples both from the vector bosons and from the charged Higgs bosons.
Note that interactions of $H_a$ and $H_b$ are easily reproduced from the interactions of $H_1$ and $H_2$ by picking the specific value of $\alpha=0$, yielding (unphysical) couplings $e_a$, $e_b$, $q_a$ and $q_b$:
\bea
e_a=v,\quad
e_b=
e_3=0,\quad
q_a=v\lambda_3,\quad
q_b=
q_3=0,\label{BCcouplingszero}
\eea
coinciding with Eq.~(\ref{Eq:IDM-couplings}), implying that Case \textoverline{BC} is physically indistinguishable from Case CC (IDM) with the additional constraint of mass degeneracy between one inert and one non-inert neutral scalar.
A similar interpretation will be applicable in other cases of mass degeneracy.

This is an example of a mass degeneracy which is not preserved by radiative corrections, which we will discuss in more detail in section \ref{sec:rge}.

\subsubsection{Spontaneously broken \texorpdfstring{$Z_2$}{Z2} symmetry}
With the fields expanded as in (\ref{vevsZ2}), the potential of (\ref{Eq:potZ2rephased}) and a vacuum of the form of (\ref{vacuum}),
setting\footnote{Other values of $\xi$ satisfying $s_{2\xi}=0$ yield similar results.} $\xi=0$,
the mass-squared matrix becomes
\bea
{\cal M}^2=
\begin{pmatrix}
	v_1^2\lambda_1   &  v_1v_2(\lambda_3+\lambda_4+\Re\lambda_5)   & 0   \\
	v_1v_2(\lambda_3+\lambda_4+\Re\lambda_5)   &  v_2^2\lambda_2  & 0   \\
	0    &  0   & -v^2\Re\lambda_5
\end{pmatrix},
\eea
and the charged mass
\bea
\mchsq=-\frac{1}{2}v^2\left(\lambda_4+\Re\lambda_5\right).
\eea
\paragraph{No mass degeneracy (RGE stable)}\mbox{}\\
Provided there is no mass degeneracy, the rotation matrix for the neutral sector is simply
\bea
R=\begin{pmatrix}
	c_\alpha    &  s_\alpha   & 0   \\
	-s_\alpha   &  c_\alpha  & 0   \\
	0    &  0   & 1
\end{pmatrix},
\label{eq:aldef}
\eea
where $\alpha$ is fixed from the diagonalization procedure. It should be emphasised that,
in distinction from that used in eq.~\eqref{eq:roa}, this angle is not arbitrary or unphysical.
We find the couplings
\bea
e_1=v_1\cos \alpha+v_2\sin\alpha,\quad
e_2=-v_1\sin\alpha+v_2\cos\alpha,\quad
e_3=0,\label{BDes}
\eea
and
\bea
q_1&=&\frac{
	v_1c_\alpha[\lambda_3v_1^2+(\lambda_1-\lambda_4-\Re\lambda_5)v_2^2]
	+v_2s_\alpha[(\lambda_2-\lambda_4-\Re\lambda_5)v_1^2+\lambda_3v_2^2]
}{v^2},\nonumber\\
q_2&=&\frac{
	v_2c_\alpha[(\lambda_2-\lambda_4-\Re\lambda_5)v_1^2+\lambda_3v_2^2]
	-v_1s_\alpha[\lambda_3v_1^2+(\lambda_1-\lambda_4-\Re\lambda_5)v_2^2]
}{v^2},\nonumber\\
q_3&=&0,\nonumber\\
q&=&\frac{1}{2v^4}[\lambda_1v_2^4+\lambda_2v_1^4+2(\lambda_3+\lambda_4+\Re\lambda_5)v_1^2v_2^2].\label{BDqs}
\eea
We find the following two identities to be satisfied in this case for $i,j=1,2$:
\bea \label{Eq:Z2-spont-br}
2(e_j^2M_i^2+e_i^2M_j^2)M_{H^\pm}^2&=&v^2(e_jq_jM_i^2+e_iq_iM_j^2-M_j^2M_i^2),\nonumber\\
2(e_j^2M_i^2+e_i^2M_j^2)q&=&(e_jq_i-e_iq_j)^2+M_i^2M_j^2.
\eea
Thus, this is a realization of Case CD.
\paragraph{Partial mass degeneracy $M_2=M_3$ (RGE unstable)}\mbox{}\\
If we allow for mass degeneracy\footnote{Mass degeneracy between $M_1$ and $M_3$ yields a similar result.} between $M_2$ and $M_3$,
the rotation matrix for the neutral sector becomes
\bea
R=\begin{pmatrix}
	1    &  0   & 0   \\
	0   &  c_\theta  & s_\theta   \\
	0    &  -s_\theta   & c_\theta
\end{pmatrix}
\begin{pmatrix}
	c_\alpha    &  s_\alpha   & 0   \\
	-s_\alpha   &  c_\alpha  & 0   \\
	0    &  0   & 1
\end{pmatrix},\label{Rthetaalpha}
\eea
where $\alpha$ is fixed by the diagonalization of the mass-squared matrix, and $\theta$ is an arbitrary angle, again an artifact resulting from the mass degeneracy.
This yields the couplings
\bea
e_1&=&v_1\cos \alpha+v_2\sin\alpha,\nonumber\\
e_2&=&-\cos\theta(v_1\sin\alpha-v_2\cos\alpha),\nonumber\\
e_3&=&\sin\theta(v_1\sin\alpha-v_2\cos\alpha).
\eea
We also find
\bea
q_1&=&\frac{
	v_1c_\alpha[\lambda_3v_1^2+(\lambda_1-\lambda_4-\Re\lambda_5)v_2^2]
	+v_2s_\alpha[(\lambda_2-\lambda_4-\Re\lambda_5)v_1^2+\lambda_3v_2^2]
}{v^2},\nonumber\\
q_2&=&c_\theta\frac{
	v_2c_\alpha[(\lambda_2-\lambda_4-\Re\lambda_5)v_1^2+\lambda_3v_2^2]
	-v_1s_\alpha[\lambda_3v_1^2+(\lambda_1-\lambda_4-\Re\lambda_5)v_2^2]
}{v^2},\nonumber\\
q_3&=&-s_\theta\frac{
	v_2c_\alpha[(\lambda_2-\lambda_4-\Re\lambda_5)v_1^2+\lambda_3v_2^2]
	-v_1s_\alpha[\lambda_3v_1^2+(\lambda_1-\lambda_4-\Re\lambda_5)v_2^2]
}{v^2},\nonumber\\
q&=&\frac{1}{2v^4}[\lambda_1v_2^4+\lambda_2v_1^4+2(\lambda_3+\lambda_4+\Re\lambda_5)v_1^2v_2^2].
\eea

We find that for any value of $\theta$, $e_2q_3-e_3q_2=0$, and the following two identities are satisfied for $i,j,k=2,3,1$:
\bea
2\left[e_k^2M_i^2+(e_i^2+e_j^2)M_k^2\right]M_{H^\pm}^2&=&v^2\left[(e_iq_i+e_jq_j)M_k^2+e_kq_kM_i^2-M_i^2M_k^2\right],\nonumber\\
2\left[e_k^2M_i^2+(e_i^2+e_j^2)M_k^2\right]q&=&(e_iq_k-e_kq_i)^2+(e_jq_k-e_kq_j)^2+M_i^2M_k^2.
\eea
Thus, this is a realization of Case \textoverline{BD}.

The situation is analogous to Case \textoverline{BC} already discussed, except that now it is $H_2$ and $H_3$ that are mass degenerate, and $\theta$ is the arbitrary unphysical angle. We adopt the same approach, defining states $H_b$ and $H_c$ (equivalent to setting $\theta=0$) resulting in couplings
\bea
e_1=v_1\cos \alpha+v_2\sin\alpha,\quad
e_b=-v_1\sin\alpha+v_2\cos\alpha,\quad
e_c=0\,,
\label{eq:es_BD}
\eea
together with
\bea
q_1&=&\frac{
	v_1c_\alpha[\lambda_3v_1^2+(\lambda_1-\lambda_4-\Re\lambda_5)v_2^2]
	+v_2s_\alpha[(\lambda_2-\lambda_4-\Re\lambda_5)v_1^2+\lambda_3v_2^2]
}{v^2},\nonumber\\
q_b&=&\frac{
	v_2c_\alpha[(\lambda_2-\lambda_4-\Re\lambda_5)v_1^2+\lambda_3v_2^2]
	-v_1s_\alpha[\lambda_3v_1^2+(\lambda_1-\lambda_4-\Re\lambda_5)v_2^2]}{v^2},\nonumber\\
q_c&=& 0.
\label{eq:qs_BD}
\eea

Comparing to (\ref{BDes}) and (\ref{BDqs}), we see that Case \textoverline{BD} is in fact physically equivalent to Case CD with the additional constraint of mass degeneracy between one CP-odd and one CP-even scalar. However, the condition for mass degeneracy,
\bea
&&(\lambda_1+\Re\lambda_5)\Re\lambda_5 v_1^4+
(\lambda_2+\Re\lambda_5)\Re\lambda_5 v_2^4\nonumber\\
&&+\left[
(\lambda_1+\Re\lambda_5)(\lambda_2+\Re\lambda_5)-(\lambda_3+\lambda_4)(\lambda_3+\lambda_4+2\Re\lambda_5)
\right]v_1^2v_2^2=0,
\eea
is not RGE stable.
\paragraph{Full mass degeneracy (RGE unstable)}\mbox{}\\
If we allow for full mass degeneracy, $M_1=M_2=M_3$, this requires $\lambda_3+\lambda_4+\Re\lambda_5=0$, $v_1^2\lambda_1=v_2^2\lambda_2$ and $v^2\Re\lambda_5=-v_1^2\lambda_1$. The mass-squared  matrix becomes
\bea
{\cal M}^2=v_1^2\lambda_1
I_3,
\eea
The rotation matrix for the neutral sector takes the generic form
\bea
R=\begin{pmatrix}
	c_1\,c_2 & s_1\,c_2 & s_2 \\
	- (c_1\,s_2\,s_3 + s_1\,c_3)
	& c_1\,c_3 - s_1\,s_2\,s_3 & c_2\,s_3 \\
	- c_1\,s_2\,c_3 + s_1\,s_3
	& - (c_1\,s_3 + s_1\,s_2\,c_3) & c_2\,c_3
\end{pmatrix},\label{fullrotationmatrix}
\eea
where $\alpha_1$, $\alpha_2$ and $\alpha_3$ are all arbitrary. In particular, we find the couplings
\bea
e_i=v_1R_{i1}+v_2R_{i2},\quad
q_i=\frac{
	(\lambda_1v_2^2+\lambda_3v^2)v_1R_{i1}+(\lambda_2v_1^2+\lambda_3v^2)v_2R_{i2}	
}{v^2},\quad
q=\frac{\lambda_1v_2^4+\lambda_2v_1^4}{2v^4}.\nonumber\\
\eea
We also find that the following two identities are satisfied
\bea
2M_{H^\pm}^2=e_1q_1+e_2q_2+e_3q_3-M_1^2,\quad
2M_1^2v^2q={\cal Q}^2+M_1^4.
\eea
for any combination of $\alpha_1$, $\alpha_2$ and $\alpha_3$.
Thus, this is a realization of Case \textoverline{AD}.

We may define states $H_a$, $H_b$ and $H_c$ (equivalent to setting $\alpha_2=\alpha_3=0)$ to regain the couplings
\bea
e_a=v_1\cos \alpha_1+v_2\sin\alpha_1,\quad
e_b=-v_1\sin\alpha_1+v_2\cos\alpha_1,\quad
e_c=0\,
\eea
\bea
q_a&=&\frac{
	(\lambda_1v_2^2+\lambda_3v^2)v_1\cos\alpha_1+(\lambda_2v_1^2+\lambda_3v^2)v_2\sin\alpha_1	
}{v^2},\nonumber\\
q_b&=&\frac{
	-(\lambda_1v_2^2+\lambda_3v^2)v_1\sin\alpha_1+(\lambda_2v_1^2+\lambda_3v^2)v_2\cos\alpha_1	
}{v^2},\nonumber\\
q_c&=&0,\nonumber\\
q&=&\frac{\lambda_1v_2^4+\lambda_2v_1^4}{2v^4},
\eea
which we recognize as the couplings of Case CD with the additional constraint of full mass degeneracy imposed\footnote{Also the angle $\alpha_1$ is arbitrary, and we may employ this arbitrariness to set, for instance, $e_a=v$ and $e_b=0$.}.
However,
the conditions needed for full mass degeneracy are not
preserved by the RGE.

With the exception of Case \textoverline{ABBB}, which will be discussed in the U(1) section, we have demonstrated that all the cases of $Z_2$ symmetry are in fact realizable in terms of explicit construction of the potential and vacuum as intended. Some of
those symmetry conditions, however, are not preserved by radiative corrections.
The only radiatively stable symmetry conditions, then, correspond to cases CC: the IDM,
with one ``active" neutral scalar and two ``dark'' ones which do not couple to gauge bosons,
and CD: a vacuum with spontaneous breaking of $Z_2$, with two CP-even and one CP-odd neutral scalar.

\subsection{U(1) symmetry}
If the potential is invariant under U(1), there exists a basis in which $m_{12}^2=0$ and $\lambda_5=\lambda_6=\lambda_7=0$.
In such a basis, it is given by
\begin{align}
V(\Phi_1,\Phi_2) &= -\frac12\left\{m_{11}^2\Phi_1^\dagger\Phi_1
+ m_{22}^2\Phi_2^\dagger\Phi_2 \right\} \nonumber \\
& + \frac{\lambda_1}{2}(\Phi_1^\dagger\Phi_1)^2
+ \frac{\lambda_2}{2}(\Phi_2^\dagger\Phi_2)^2
+ \lambda_3(\Phi_1^\dagger\Phi_1)(\Phi_2^\dagger\Phi_2)
+ \lambda_4(\Phi_1^\dagger\Phi_2)(\Phi_2^\dagger\Phi_1).\label{Eq:potU1}
\end{align}
Again, we parametrize the Higgs doublets in the most general way, following the steps in Eqs. (\ref{vevsZ2}) through (\ref{orthogonal}).
Without loss of generality we can independently rephase $\Phi_1$ and $\Phi_2$ in order to make both $\left<\Phi_1\right>$ and $\left<\Phi_2\right>$ real and non-negative. Minimizing the potential with respect to the fields yields the stationary-point equations
\bea
v_1\left[-m_{11}^2+v_1^2\lambda_1+v_2^2(\lambda_3+\lambda_4)\right]=0,\quad
v_2\left[-m_{22}^2+v_2^2\lambda_2+v_1^2(\lambda_3+\lambda_4)\right]=0.
\eea
Allowing for solutions with one vanishing VEV, the stationary-point equations are solved by simply putting
\bea
v_2=0,\quad
m_{11}^2=v^2\lambda_1.
\eea
The $v_1=0$ case is related to the $v_2=0$ case by simply interchanging the two doublets.

If both $v_i\neq0$, then the stationary-point equations are solved by
\bea
m_{11}^2=v_1^2\lambda_1+v_2^2(\lambda_3+\lambda_4),\quad
m_{22}^2=v_2^2\lambda_2+v_1^2(\lambda_3+\lambda_4).
\label{eq:u1m}
\eea
We see that these two ways of satisfying the stationary-point equations are topologically different, meaning one cannot get from the second solution to the first one by simply letting $v_2\to0$ in a continuous way. We would therefore expect these two situations to lead to different physics.
\subsubsection{U(1) symmetric potential and vacuum}
\label{sec:bcc}
With the potential of (\ref{Eq:potU1}) and a vacuum of the form
\bea
\left<\Phi_1\right>=\frac{1}{\sqrt{2}}\left(
\begin{array}{c}0\\ v
\end{array}\right),\quad
\left<\Phi_2\right>=\left(
\begin{array}{c}0\\ 0
\end{array}\right),
\eea
the mass-squared matrix becomes
\bea
{\cal M}^2=v^2\text{diag}\left[\lambda_1,\half\left(\lambda_3+\lambda_4-\frac{m_{22}^2}{v^2}\right),\half\left(\lambda_3+\lambda_4-\frac{m_{22}^2}{v^2}\right)\right],
\eea
and the charged mass
\bea
\mchsq=\frac{1}{2}\left(v^2\lambda_3-m_{22}^2\right).
\eea
Note that this vacuum does not break the U(1) symmetry defined by (\ref{u1def}) -- when the $\Phi_1$ doublet receives a phase as a result of the U(1), that phase could be absorbed via simultaneous (global) rephasing of both doublets, which is always allowed due to the hypercharge gauge symmetry. Therefore effectively the vacuum would remain invariant.
\paragraph{Partial mass degeneracy $M_2=M_3$ (RGE stable)}\mbox{}\\
We see from the mass matrix that the U(1) symmetry dictates mass degeneracy between two of the neutral scalars.
Provided there is no full mass degeneracy, we obtain the physical couplings of the neutral
scalars to electroweak gauge bosons and charged scalars\footnote{Note that due to the mass degeneracy, we may put in an arbitrary rotation angle in the rotation matrix of the neutral sector. However, none of the couplings or masses presented here actually depends on this (unphysical) arbitrary angle.},
\bea
e_1=v,\quad
e_2=e_3=0,\quad
\label{eq:c1}
q_1=v\lambda_3,\quad
q_2=q_3=0.
\label{eq:c2}
\eea
This is a realization of Case BCC, which is a version of the IDM with a Peccei--Quinn
symmetry instead of the $Z_2$ one. This model predicts degenerate dark matter candidates, and is
disfavoured by astronomical observations~\cite{LopezHonorez:2006}. The mass degeneracy is here of a different kind than what we encountered when discussing $Z_2$-symmetric cases, since now both states are inert. While the cases of mass degenerate states discussed for $Z_2$ are radiatively unstable, the mass degenerate states encountered here are radiatively stable under the RGE, as will be discussed in section~\ref{sec:rge}.
\paragraph{Full mass degeneracy (RGE unstable)}\mbox{}\\
Allowing for full mass degeneracy requires $m_{22}^2=(-2\lambda_1+\lambda_3+\lambda_4)v^2$.
The mass-squared matrix then becomes
\bea
{\cal M}^2=v^2 \lambda_1
I_3,
\eea
and the charged mass
\bea
\mchsq=\frac{v^2}{2}\left(2\lambda_1-\lambda_4\right).
\eea
Due to the full mass degeneracy, the rotation matrix for the neutral sector is now given by the most general form (\ref{fullrotationmatrix}), yielding couplings
	\bea
	e_i=vR_{i1},\quad
	q_i=\lambda_3vR_{i1}.
	\eea
We find that ${\cal Q}^2=0$ for any combination of $\alpha_1$, $\alpha_2$ and $\alpha_3$.
Thus, this is a realization of Case \textoverline{ABBB}.
	
Like before,
we may define states $H_a$, $H_b$ and $H_c$ (equivalent to setting $\alpha_1=\alpha_2=\alpha_3=0)$ to regain the couplings
\bea
e_1=v,\quad
e_2=e_3=0,\quad
q_1=v\lambda_3,\quad
q_2=q_3=0.
\eea
which we recognize as the couplings of Case BCC with the additional constraint of full mass degeneracy imposed. Thus, we see that Case \textoverline{ABBB} is equivalent to Case BCC with the additional constraint of full mass degeneracy imposed.
The full mass-degeneracy constraint is unstable under RGE, see section \ref{sec:rge}.
\subsubsection{Spontaneously broken U(1) symmetry}
\label{sec:su1}
With the potential of (\ref{Eq:potU1}) and a vacuum of the form
\bea
\left<\Phi_1\right>=\frac{1}{\sqrt{2}}\left(
\begin{array}{c}0\\ v_1
\end{array}\right),\quad
\left<\Phi_2\right>=\frac{1}{\sqrt{2}}\left(
\begin{array}{c}0\\ v_2
\end{array}\right),
\eea
the mass-squared matrix becomes
\bea
{\cal M}^2=
\begin{pmatrix}
	v_1^2\lambda_1   &  v_1v_2(\lambda_3+\lambda_4)   & 0   \\
	v_1v_2(\lambda_3+\lambda_4)   &  v_2^2\lambda_2  & 0   \\
	0    &  0   & 0
\end{pmatrix},
\label{eq:mm}
\eea
with $M_3=0$, and the charged mass
\bea
\mchsq=-\frac{1}{2}v^2\lambda_4.
\eea
\paragraph{No mass degeneracy (RGE stable)}\mbox{}\\
Provided there is no mass degeneracy, the rotation matrix for the neutral sector is simply
\bea
R=\begin{pmatrix}
	c_\alpha    &  s_\alpha   & 0   \\
	-s_\alpha   &  c_\alpha  & 0   \\
	0    &  0   & 1
\end{pmatrix},
\eea
where $\alpha$ is fixed by the diagonalization of the mass matrix.
The resulting couplings are
\bea
e_1=v_1\cos \alpha+v_2\sin\alpha,\quad
e_2=-v_1\sin\alpha+v_2\cos\alpha,\quad
e_3=0,\label{u1e}
\eea
and
\bea
q_1&=&\frac{
	v_1c_\alpha[\lambda_3v_1^2+(\lambda_1-\lambda_4)v_2^2]
	+v_2s_\alpha[(\lambda_2-\lambda_4)v_1^2+\lambda_3v_2^2]
}{v^2},\nonumber\\
q_2&=&\frac{
	v_2c_\alpha[(\lambda_2-\lambda_4)v_1^2+\lambda_3v_2^2]
	-v_1s_\alpha[\lambda_3v_1^2+(\lambda_1-\lambda_4)v_2^2]
}{v^2},\nonumber\\
q_3&=&0,\nonumber\\
q&=&\frac{1}{2v^4}[\lambda_1v_2^4+\lambda_2v_1^4+2(\lambda_3+\lambda_4)v_1^2v_2^2].\label{u1q}
\eea
We find the following two identities to be satisfied in this case for $i,j=1,2$:
\bea
2(e_j^2M_i^2+e_i^2M_j^2)M_{H^\pm}^2&=&v^2(e_jq_jM_i^2+e_iq_iM_j^2-M_j^2M_i^2),\nonumber\\
2(e_j^2M_i^2+e_i^2M_j^2)q&=&(e_jq_i-e_iq_j)^2+M_i^2M_j^2.
\eea
Thus, this is a realization of Case C$_0$D. This is the Peccei--Quinn model~\cite{Peccei:1977hh}
in which a massless axion appears as a result of spontaneous breaking of the continuous U(1) symmetry.
The introduction of a soft breaking term in the potential prevents the masslessness, however we will not
discuss soft breaking of symmetries here.

Another possibility, however, is to promote the U(1)
symmetry to a local gauge symmetry, thus introducing a new gauge boson, $Z^\prime$
(see, for instance,~\cite{Fukuda:2017ylt,Campos:2017dgc,Camargo:2018uzw} and references therein). The massless scalar
resulting from the spontaneous U(1) breaking is then responsible for giving $Z^\prime$ its
mass, and we are left with a scalar sector including a charged scalar and two CP-even scalars,
which is contained in the mass spectrum predicted for Case C$_0$D.  This scalar sector
can therefore be of  phenomenological interest, even without a soft symmetry breaking parameter.

\paragraph{Partial mass degeneracy $M_2=M_3=0$ (RGE unstable)}\mbox{}\\
Allowing for mass degeneracy\footnote{Mass degeneracy between $M_1$ and $M_3$ yields a similar result.} 
$M_2=M_3=0$\footnote{\label{footnote:degeneracy}The existence of massless scalars that are not Goldstone bosons enables the
emergence of non-zero vacuum expectation values generated radiatively, see \cite{Georgi:1974au}. 
We will not discuss this unphysical case of zero-mass degeneracy.}, the rotation matrix is again given as in (\ref{Rthetaalpha})
where $\alpha$ is fixed by the diagonalization of the mass matrix, and $\theta$ is an arbitrary angle. Working out the couplings, we find that $e_2q_3-e_3q_2=0$,
 and the following two identities are satisfied for $i,j,k=2,3,1$:
\bea
2(e_i^2+e_j^2)M_{H^\pm}^2&=&v^2(e_iq_i+e_jq_j),\nonumber\\
2(e_i^2+e_j^2)M_k^2q&=&(e_iq_k-e_kq_i)^2+(e_jq_k-e_kq_j)^2,
\eea
for all values of $\theta$.
Thus, this is a realization of Case \textoverline{B$_0$D}.

Like before, we may define states $H_b$ and $H_c$ (equivalent to setting $\theta=0$) and regain the couplings of (\ref{u1e}) and (\ref{u1q}).
Thus, we conclude that Case \textoverline{B$_0$D} is physically equivalent to Case C$_0$D with the additional mass degeneracy $M_2=M_3=0$.
The RGE unstable condition leading to the mass degeneracy is in this case given by $\lambda_1 \lambda_2 = (\lambda_3 + \lambda_4)^2$. We will not discuss the full mass degeneracy case where all three masses vanish here, since this implies an SO(3) symmetric model which will be treated later.
\subsection{CP2 symmetry}
\label{sec:CP2}
If the potential is invariant under CP2, there exists a basis in which $m_{12}^2=0$, $m_{22}^2=m_{11}^2$, $\lambda_2=\lambda_1$ and $\lambda_7=-\lambda_6$.
In a basis in which the CP2 symmetry is manifest, the potential is given by
\begin{align}
\label{Eq:potCP2}
V(\Phi_1,\Phi_2) &= -\frac{m_{11}^2}{2}\left\{\Phi_1^\dagger\Phi_1
+\Phi_2^\dagger\Phi_2 \right\} \nonumber \\
& + \frac{\lambda_1}{2}\left\{(\Phi_1^\dagger\Phi_1)^2
+ (\Phi_2^\dagger\Phi_2)^2\right\}
+ \lambda_3(\Phi_1^\dagger\Phi_1)(\Phi_2^\dagger\Phi_2)
+ \lambda_4(\Phi_1^\dagger\Phi_2)(\Phi_2^\dagger\Phi_1)\nonumber \\
&+ \frac12\left[\lambda_5(\Phi_1^\dagger\Phi_2)^2 + \hc\right]
+\left\{\lambda_6\left[(\Phi_1^\dagger\Phi_1)-
(\Phi_2^\dagger\Phi_2)\right](\Phi_1^\dagger\Phi_2)
+{\rm \hc}\right\}.
\end{align}
Davidson and Haber \cite{Davidson:2005cw} have demonstrated that for this potential one can change basis in order to get a similar potential, but with $\Im\lambda_5=\lambda_6=0$. We shall employ this change of basis in order to simplify the analysis.
Again, we parametrize the Higgs doublets in the most general way, following the steps in Eqs.~(\ref{vevsZ2}) through (\ref{orthogonal}).
Minimizing the potential with respect to the fields yields the stationary-point equations
\bea
v_1[-m_{11}^2+v_1^2\lambda_1+v_2^2(\lambda_3+\lambda_4+c_{2\xi}\Re\lambda_5)]&=&0,\nonumber\\
v_2[-m_{11}^2+v_2^2\lambda_1+v_1^2(\lambda_3+\lambda_4+c_{2\xi}\Re\lambda_5)]&=&0,\nonumber\\
v_1v_2s_{2\xi}\Re\lambda_5&=&0.
\eea
We will not allow for $\Re\lambda_5=0$, since this will lead to the CP3 symmetry, which we will study later. Allowing for solutions with one vanishing VEV, the stationary-point equations are solved by putting
\bea
v_2=0,\quad
m_{11}^2=v^2\lambda_1.
\eea
With a vanishing VEV, the phase $\xi$ can now be absorbed into the field $\Phi_2$.

If the two VEVs are identical, the stationary-point equations are solved by
\bea
v_2=v_1,\quad
m_{11}^2=(\lambda_1+\lambda_3+\lambda_4+c_{2\xi}\Re\lambda_5)v_1^2,\quad
s_{2\xi}=0.
\eea

If both $v_i\neq0$ and $v_2\neq v_1$, then the stationary-point equations are solved by
\bea
m_{11}^2=v^2\lambda_1,\quad
c_{2\xi}\Re\lambda_5=\lambda_1-\lambda_3-\lambda_4,\quad
s_{2\xi}=0.
\eea
This latter option leads to a CP3 conserving model, and will be studied in section~\ref{sec:cp3-symmetry}.

\subsubsection{CP2 symmetric potential with \texorpdfstring{$v_2=0$}{v2=0}}
With the potential
\begin{align}
V(\Phi_1,\Phi_2) &= -\frac{m_{11}^2}{2}\left\{\Phi_1^\dagger\Phi_1
+\Phi_2^\dagger\Phi_2 \right\} \nonumber \\
& + \frac{\lambda_1}{2}\left\{(\Phi_1^\dagger\Phi_1)^2
+ (\Phi_2^\dagger\Phi_2)^2\right\}
+ \lambda_3(\Phi_1^\dagger\Phi_1)(\Phi_2^\dagger\Phi_2)
+ \lambda_4(\Phi_1^\dagger\Phi_2)(\Phi_2^\dagger\Phi_1)\nonumber \\
&+ \frac{\Re\lambda_5}{2}\left[(\Phi_1^\dagger\Phi_2)^2 + (\Phi_2^\dagger\Phi_1)^2\right]
.\label{Eq:potCP2new}
\end{align}
and the vacuum
\bea
\left<\Phi_1\right>=\frac{1}{\sqrt{2}}\left(
\begin{array}{c}0\\ v
\end{array}\right),\quad
\left<\Phi_2\right>=\left(
\begin{array}{c}0\\ 0
\end{array}\right),
\eea
the mass-squared matrix becomes
\bea
{\cal M}^2=v^2\text{diag}(\lambda_1,\half\left[-\lambda_1+\lambda_3+\lambda_4+\Re\lambda_5\right],\half\left[-\lambda_1+\lambda_3+\lambda_4-\Re\lambda_5\right]),
\eea
and the charged mass
\bea
\mchsq=\frac{v^2}{2}\left(\lambda_3-\lambda_1\right).
\eea
\paragraph{No mass degeneracy (RGE stable)}\mbox{}\\
Provided there is no mass degeneracy, we obtain the couplings
\bea
e_1=v,\quad
e_2=e_3=0,\quad
q_1=v\lambda_3,\quad
q_2=q_3=0,\quad
q=\half{\lambda_1}.
\eea
We also find
\bea
2M_{H^\pm}^2=e_1q_1-M_1^2,\quad
2v^2q=M_1^2.
\eea
This is a realization of Case CCD, the default implementation of the CP2 model.
\paragraph{Partial mass degeneracy $M_1=M_2$ (RGE unstable)}\mbox{}\\
Allowing for mass degeneracy\footnote{Mass degeneracy between $M_1$ and $M_3$ yields a similar result.} between $M_1$ and $M_2$  requires $\Re\lambda_5=(3\lambda_1-\lambda_3-\lambda_4)$. The mass-squared matrix then becomes
\bea
{\cal M}^2=v^2\text{diag}(\lambda_1,\lambda_1,-2\lambda_1+\lambda_3+\lambda_4).
\eea
The rotation matrix is given as in (\ref{eq:roa})
where $\alpha$ is an arbitrary angle due to the mass degeneracy.
Working out the couplings, we find that
$e_1q_2-e_2q_1=0$, and that
\bea
2M_{H^\pm}^2=e_1q_1+e_2q_2-M_1^2,\quad
2v^2q=M_1^2,
\eea
for any value of $\alpha$.
Therefore this is a realization of Case \textoverline{BCD}.

Again we may define states $H_a$ and $H_b$ (equivalent to letting $\alpha=0$), with couplings given by
\begin{subequations}
\bea
e_a=v,\quad
e_b=
e_3=0,\quad
q_a=v\lambda_3,\quad
q_b=
q_3=0.
\eea
\end{subequations}
that is, we obtain two inert states, one of which is CP-odd. Thus, we interpret Case \textoverline{BCD} as physically equivalent to Case CCD with the additional mass degeneracy between one inert and one non-inert neutral scalar.
The constraint yielding mass degeneracy given above is RGE unstable, as will be discussed in
section \ref{sec:rge}.
We will not discuss mass degeneracy between $M_2$ and $M_3$ or full mass degeneracy yet, since this will result in a CP3 symmetric model which we will study later.

\subsubsection{CP2 symmetric potential with \texorpdfstring{$v_2=v_1$}{v2=v1}}
Analyzing the solution of the stationary-point equations with $v_2=v_1$ again leads to Case CCD which we already encountered, implying that this is simply a description of the same physical model in another basis. We omit the details.
\subsection{CP3 symmetry}
\label{sec:cp3-symmetry}
If the potential is invariant under CP3, there exists a basis in which $m_{12}^2=0$, $\lambda_6=\lambda_7=0$, $m_{22}^2=m_{11}^2$, $\lambda_2=\lambda_1$ and $\lambda_5=\lambda_1-\lambda_3-\lambda_4$
(real)~\cite{Ferreira:2010bm,Branco:2011iw}.
Writing out the potential in such a basis, in which the CP3 symmetry is manifest, we obtain
\begin{align}
V(\Phi_1,\Phi_2) &= -\frac{m_{11}^2}{2}\left\{\Phi_1^\dagger\Phi_1
+\Phi_2^\dagger\Phi_2 \right\} \nonumber \\
& + \frac{\lambda_1}{2}\left\{(\Phi_1^\dagger\Phi_1)^2
+ (\Phi_2^\dagger\Phi_2)^2\right\}
+ \lambda_3(\Phi_1^\dagger\Phi_1)(\Phi_2^\dagger\Phi_2)
+ \lambda_4(\Phi_1^\dagger\Phi_2)(\Phi_2^\dagger\Phi_1)\nonumber \\
&+ \frac{\lambda_1-\lambda_3-\lambda_4}{2}\left[(\Phi_1^\dagger\Phi_2)^2 + (\Phi_2^\dagger\Phi_1)^2\right]
.\label{Eq:potCP3}
\end{align}
Again, we parametrize the Higgs doublets in the most general way, following the steps in Eqs.~(\ref{vevsZ2}) through (\ref{orthogonal}).
Minimizing the potential with respect to the fields yields the stationary-point equations
\bea
v_1[-m_{11}^2+v_1^2\lambda_1+v_2^2(\lambda_3+\lambda_4+c_{2\xi}(\lambda_1-\lambda_3-\lambda_4))]&=&0,\nonumber\\
v_2[-m_{11}^2+v_2^2\lambda_1+v_1^2(\lambda_3+\lambda_4+c_{2\xi}(\lambda_1-\lambda_3-\lambda_4))]&=&0,\nonumber\\
v_1v_2s_{2\xi}(\lambda_1-\lambda_3-\lambda_4)&=&0.
\eea
We will not allow for $\lambda_1-\lambda_3-\lambda_4=0$, since this will be studied in the section on SO(3) symmetry. Allowing for solutions with one vanishing VEV, the stationary-point equations are solved by putting
\bea
v_2=0,\quad
m_{11}^2=v^2\lambda_1.
\eea
With a vanishing VEV, the phase $\xi$ can now be absorbed into the field $\Phi_2$.

If the VEVs are identical, the stationary-point conditions are satisfied whenever
\bea
v_2=v_1,\quad
m_{11}^2=2v_1^2\lambda_1,\quad
s_{\xi}=0,
\eea
or whenever
\bea
v_2=v_1,\quad
m_{11}^2=2(\lambda_3+\lambda_4)v_1^2,\quad
c_{\xi}=0.
\eea

If both $v_i\neq0$ and $v_2\neq v_1$, the stationary point conditions are satisfied whenever
\bea
\lambda_4=\lambda_1-\lambda_3,\quad
m_{11}^2=\lambda_1v^2,\quad
s_{2\xi}=0,
\eea
which leads to an SO(3)-symmetric potential, which will be discussed in section~\ref{sect:section-SO3}.
The CP3 model of reference~\cite{Ferreira:2010bm} had non-zero VEVS $v_1\neq v_2$,
but it included a soft breaking term, thus it is outside of the scope of the present work.

\subsubsection{CP3 symmetric potential with \texorpdfstring{$v_2=0$}{v2=0}}

With the potential of (\ref{Eq:potCP3}) and the vacuum
\bea
\left<\Phi_1\right>=\frac{1}{\sqrt{2}}\left(
\begin{array}{c}0\\ v
\end{array}\right),\quad
\left<\Phi_2\right>=\left(
\begin{array}{c}0\\ 0
\end{array}\right),
\eea
the mass-squared matrix becomes\footnote{CP3 is broken spontaneously in this case, so a massless Goldstone boson appears.}
\bea
{\cal M}^2=v^2\text{diag}(\lambda_1,0,-\lambda_1+\lambda_3+\lambda_4),
\eea
and the charged mass
\bea
\mchsq=\frac{v^2}{2}\left(\lambda_3-\lambda_1\right).
\eea
\paragraph{No mass degeneracy (RGE stable)}\mbox{}\\
The mass-squared matrix is diagonal, so the rotation matrix is simply the identity matrix. The couplings are found to be
\bea
e_1=v,\quad
e_2=e_3=0,\quad
q_1=v\lambda_3,\quad
q_2=q_3=0,\quad
q=\frac{\lambda_1}{2}.\label{eq:C0CD-couplings}
\eea
We also find
\bea
2M_{H^\pm}^2=e_1q_1-M_1^2,\quad
2v^2q=M_1^2.
\eea
This is a realization of Case C$_0$CD.
\paragraph{Partial mass degeneracy $M_1=M_2=0$ (RGE unstable)}\mbox{}\\
Allowing for mass degeneracy between $M_1$ and $M_2$ requires $\lambda_1=0$.
The mass-squared matrix then becomes (see footnote~\ref{footnote:degeneracy})
\bea
{\cal M}^2=v^2(\lambda_3+\lambda_4)\text{diag}(0,0,1),
\eea
and the charged mass
\bea
\mchsq=\frac{1}{2}v^2\lambda_3.
\eea
The rotation matrix is given as in (\ref{eq:roa})
where $\alpha$ is an arbitrary angle due to the mass degeneracy.
Working out the couplings, we find that
$e_1q_2-e_2q_1=0$, $2M_{H^\pm}^2=e_1q_1+e_2q_2$ and $q=0$ for any value of $\alpha$. Therefore, this is a realization of Case \textoverline{B$_0$CD}.

Like before, we may define states $H_a$ and $H_b$ such that
\begin{subequations} \label{Eq:B0CD-couplings}
\bea
e_a=v,\quad
e_b=
e_3=0,\quad
q_a=v\lambda_3,\quad
q_b=
q_3=0.
\eea
\end{subequations}
Thus, we interpret Case \textoverline{B$_0$CD} as physically equivalent to Case C$_0$CD with the mass degeneracy $M_1=M_2=0$ imposed in addition.
Again, the condition responsible for the mass degeneracy is RGE unstable, as will be seen in section~\ref{sec:rge}.

\paragraph{Partial mass degeneracy $M_1=M_3$ (RGE unstable)}\mbox{}\\
Allowing for mass degeneracy between $M_1$ and $M_3$ yields $\lambda_4=2\lambda_1-\lambda_3$. The mass-squared matrix then becomes
\bea
{\cal M}^2=v^2 \lambda_1
\text{diag}\{1,0,1\},
\eea
and the charged mass
\bea
\mchsq=\frac{1}{2}v^2(\lambda_3-\lambda_1).
\eea
We write out the most general rotation matrix\footnote{$R=\begin{pmatrix}
		c_\alpha    &  0   & s_\alpha   \\
		0   &  0  & 0   \\
		-s_\alpha    &  0   & c_\alpha
	\end{pmatrix}$, where $\alpha$ is arbitrary.} compatible with the mass degeneracy and work out the couplings to find
	 $e_1q_3-e_3q_1=0$ and $2M_{H^\pm}^2=e_1q_1+e_3q_3-M_1^2$ and $2v^2q=M_1^2$ for any value of $\alpha$.
This is a realization of Case \textoverline{BC$_0$D}, in which one would have a degenerate pair, one of which is CP odd, together with a massless CP-even scalar.

Defining states $H_a$ and $H_c$, equivalent to putting $\alpha=0$, we get
\bea
e_a=v,\quad
e_2=e_c=0,\quad
q_a=v\lambda_3,\quad
q_2=q_c=0.
\eea
Thus, we interpret Case \textoverline{BC$_0$D} as physically equivalent to Case C$_0$CD with the mass degeneracy $M_1=M_3$ imposed in addition.
The condition responsible for the mass degeneracy is however RGE unstable.
We will not yet discuss mass degeneracy between $M_2$ and $M_3$ or full mass degeneracy, since this will lead to an SO(3) symmetric model which will be treated in section~\ref{sect:section-SO3}.

\subsubsection{CP3 symmetric potential with \texorpdfstring{$v_2=v_1$}{v2=v1} and \texorpdfstring{$s_\xi=0$}{sin xi=0}}
Analyzing the solution of the stationary-point equations with $v_2=v_1$ again leads to Case C$_0$CD which we already encountered, implying that this is simply a description of the same physical model in another basis. We omit the details.

\subsubsection{CP3 symmetric potential with \texorpdfstring{$v_2=v_1$}{v2=v2} and \texorpdfstring{$c_\xi=0$}{cos xi=0}}
With the potential of (\ref{Eq:potCP3}) and the vacuum
\bea
\left<\Phi_1\right>=\frac{1}{\sqrt{2}}\left(
\begin{array}{c}0\\ v_1
\end{array}\right),\quad
\left<\Phi_2\right>=\frac{e^{i\xi}}{\sqrt{2}}\left(
\begin{array}{c}0\\ v_1
\end{array}\right),
\eea
putting\footnote{Putting $\xi=-\pi/2$ yields a similar result.} $\xi=\pi/2$, the squared mass matrix becomes
\bea
{\cal M}^2=v_1^2
\begin{pmatrix}
	\lambda_1   &  -\lambda_1+2\lambda_3+2\lambda_4   & 0   \\
	-\lambda_1+2\lambda_3+2\lambda_4   &  \lambda_1  & 0   \\
	0    &  0   & 2(\lambda_1-\lambda_3-\lambda_4)
\end{pmatrix},
\eea
and the charged mass
\bea
\mchsq=v_1^2(\lambda_1-\lambda_3-2\lambda_4).
\eea
\paragraph{Partial mass degeneracy $M_2=M_3$ (RGE stable)}\mbox{}\\
We have two mass degenerate scalars, and the mass matrix is diagonalized by
\bea
R=
\begin{pmatrix}
	1    &  0   & 0   \\
	0   &  c_\alpha  & s_\alpha   \\
	0    &  -s_\alpha   & c_\alpha
\end{pmatrix}
\begin{pmatrix}
	\frac{1}{\sqrt{2}}    &  \frac{1}{\sqrt{2}}   & 0   \\
	-\frac{1}{\sqrt{2}}   &  \frac{1}{\sqrt{2}}  & 0   \\
	0    &  0   & 1
\end{pmatrix},
\eea
where $\alpha$ is arbitrary due to the mass degeneracy. However, none of the masses or couplings depend on $\alpha$. We find
 the couplings
\bea
e_1=v,\quad
e_2=e_3=0,\quad
q_1=v(\lambda_1-\lambda_4),\quad
q_2=q_3=0,\quad
q=\frac{1}{2}(\lambda_3+\lambda_4),
\eea
and
\bea
2M_{H^\pm}^2=e_1q_1-M_1^2,\quad
2v^2q=M_1^2.
\eea
This is a realization of Case BCCD.
\paragraph{Full mass degeneracy (RGE unstable)}\mbox{}\\
Allowing for full mass degeneracy requires $\lambda_4=\half(\lambda_1-2\lambda_3)$,
The neutral-sector mass-squared matrix becomes
\bea
{\cal M}^2=v_1^2\lambda_1
I_3,
\eea
and the charged mass
\bea
\mchsq=v_1^2\lambda_3.
\eea
The rotation matrix for the neutral sector is given by (\ref{fullrotationmatrix}) with arbitrary $\alpha_1$, $\alpha_2$ and $\alpha_3$. Now we find
\bea
e_i=v_1(R_{i1}+R_{i2}),\quad
q_i=\half(\lambda_1+2\lambda_3)v_1(R_{i1}+R_{i2}),\quad
q=\frac{1}{4}\lambda_1.
\eea
We also find that ${\cal Q}^2=0$ and
\bea
2M_{H^\pm}^2=e_1q_1+e_2q_2+e_3q_3-M_1^2,\quad
2v^2q=M_1^2
\eea
for any values of $\alpha_1$, $\alpha_2$ and $\alpha_3$.
Thus, this is a realization of Case \textoverline{ABBBD}.

Like before,
we may define states $H_a$, $H_b$ and $H_c$ (equivalent to setting $\alpha_1=\pi/4$ and $\alpha_2=0$) to regain the couplings
\bea
e_1=v,\quad
e_2=e_3=0,\quad
q_1=v(\lambda_1-\lambda_4),\quad
q_2=q_3=0,\quad
q=\frac{1}{2}(\lambda_3+\lambda_4),
\eea
which we recognize as the couplings of Case BCCD with the additional constraint of full mass degeneracy imposed. Thus, we se that Case \textoverline{ABBBD} is Case BCCD with the additional constraint of full mass degeneracy imposed.
Once more, the condition responsible for the full mass degeneracy is RGE unstable.
\subsection{SO(3) symmetry}
\label{sect:section-SO3}
If the potential is invariant under SO(3), there exists a basis in which $m_{12}^2=0$, $\lambda_5=\lambda_6=\lambda_7=0$, $m_{22}^2=m_{11}^2$, $\lambda_2=\lambda_1$ and $\lambda_4=\lambda_1-\lambda_3$~\cite{Ferreira:2010yh,Branco:2011iw}.
Writing out the potential in such a basis, it is given by
\begin{align}
V(\Phi_1,\Phi_2) &= -\frac{m_{11}^2}{2}\left\{\Phi_1^\dagger\Phi_1
+\Phi_2^\dagger\Phi_2 \right\}
+ \frac{\lambda_1}{2}\left\{(\Phi_1^\dagger\Phi_1)^2
+ (\Phi_2^\dagger\Phi_2)^2\right\} \nonumber \\
& + \lambda_3(\Phi_1^\dagger\Phi_1)(\Phi_2^\dagger\Phi_2)
+ (\lambda_1-\lambda_3)(\Phi_1^\dagger\Phi_2)(\Phi_2^\dagger\Phi_1)
.\label{Eq:potSO3}
\end{align}
Minimizing the potential with respect to the fields yields the stationary-point equations
\bea
v_1(-m_{11}^2+v^2\lambda_1)=0,\quad
v_2(-m_{11}^2+v^2\lambda_1)=0,
\eea
solved whenever $m_{11}^2=v^2\lambda_1$. The parameters of the SO(3)-invariant potential are insensitive to basis changes, and therefore we choose to do the analysis by working in the Higgs basis.
With the potential of (\ref{Eq:potSO3}) and the vacuum
\bea
\left<\Phi_1\right>=\frac{1}{\sqrt{2}}\left(
\begin{array}{c}0\\ v
\end{array}\right),\quad
\left<\Phi_2\right>=\left(
\begin{array}{c}0\\ 0
\end{array}\right),
\eea
the mass-squared matrix becomes
\bea
{\cal M}^2=v^2\lambda_1
\text{diag}\{1,0,0\},
\eea
and the charged mass
\bea
\mchsq=\frac{v^2}{2}\left(\lambda_3-\lambda_1\right).
\eea
The vacuum breaks U(2) so that there are 2 Goldstone bosons.
\paragraph{Partial mass degeneracy $M_2=M_3=0$ (RGE stable)}\mbox{}\\
Again, there is arbitrariness in the rotation matrix\footnote{
	$R=\begin{pmatrix}
		1    &  0   & 0   \\
		0   &  c_\alpha  & s_\alpha   \\
		0    &  -s_\alpha   & c_\alpha
	\end{pmatrix}$,
	with $\alpha$ being arbitrary.} due to the two mass degenerate states. The couplings are found to be
\bea
e_1=v,\quad
e_2=e_3=0,\quad
q_1=v\lambda_3,\quad
q_2=q_3=0,\quad
q=\frac{\lambda_1}{2}.
\eea
We also find
\bea
2M_{H^\pm}^2=e_1q_1-M_1^2,\quad
2v^2q=M_1^2.
\eea
This is a realization of Case B$_0$C$_0$C$_0$D.
\paragraph{Full mass degeneracy (RGE unstable)}\mbox{}\\
Allowing for full mass degeneracy requires $\lambda_1=0$.
The neutral-sector mass-squared matrix becomes the three-by-three zero matrix,
and the charged mass
\bea
\mchsq=\half v^2\lambda_3.
\eea
The rotation matrix for the neutral sector is given by (\ref{fullrotationmatrix}) with arbitrary $\alpha_1$, $\alpha_2$ and $\alpha_3$. Now we find
\bea
e_i=vR_{i1},\quad
q_i=\lambda_3vR_{i1},\quad
q=0.
\eea
We also find that ${\cal Q}^2=0$, and
\bea
2M_{H^\pm}^2&=&e_1q_1+e_2q_2+e_3q_3-M_1^2,
\eea
for any value of $\alpha_1$, $\alpha_2$ and $\alpha_3$.
Thus, this is a realization of Case \textoverline{A$_0$B$_0$B$_0$B$_0$D}.

Like before,
we may define states $H_a$, $H_b$ and $H_c$ (equivalent to setting $\alpha_1=\alpha_2=\alpha_3=0$) to regain the couplings
\bea
e_1=v,\quad
e_2=e_3=0,\quad
q_1=v\lambda_3,\quad
q_2=q_3=0.
\eea
Thus, we may interpret Case \textoverline{A$_0$B$_0$B$_0$B$_0$D} as physically equivalent to Case B$_0$C$_0$C$_0$D with the additional constraint of full mass degeneracy imposed.
For this case, mass degeneracy is also unstable under radiative corrections.

\section{RGE stability of symmetry conditions}
\label{sec:rge}
\setcounter{equation}{0}

In the previous section we found several cases
of 2HDM symmetries which required specific conditions imposed on physical parameters specifying the potential.
Whenever conditions defining the cases are satisfied, the scalar sector (i.e., the scalar potential)
is invariant under a given symmetry, e.g. CP1, $Z_2$, etc. For each symmetry the corresponding list of cases
is complete in the sense that there exists no other case compatible with invariance of the potential under the considered symmetry.
That suggests a natural classification of the cases:
\bit
\item cases which are stable under the RGE,
\item cases which are unstable under the RGE,
\eit
where stability is defined by the vanishing of the perturbative beta functions corresponding to all conditions specifying a
given case\footnote{Usually a case is defined by several relations among observables like couplings or masses.}.
{\it Importantly, the perturbative expansion can not violate the symmetry\footnote{We are not considering here symmetries which are anomalous.},
therefore after including radiative corrections the potential is still symmetric and, due to the completeness, we must remain within one of the cases available to the considered symmetry.}
If the case is preserved by radiative corrections, i.e., the relation between couplings specific to that case is RGE invariant, the case is stable.
Otherwise, radiative corrections force us to move to another case, that happens when the beta function
of a given condition is non-zero.
The lesson is that it may happen (in unstable cases) that conditions crucial for the presence of a symmetry may be violated by radiative corrections, even though, still, the symmetry is preserved. The starting case would be replaced by some other case. This is an unfamiliar consequence of a symmetry.

In other words, the cases unstable under loop corrections would correspond to tree-level fine tunings
that are violated when radiative corrections are taken into account. The remaining, stable, cases constitute
constraints on the parameters of the model which are preserved under renormalization,
even when spontaneous (or soft) symmetry breaking is involved.

A simple way to analyze this is
to use $\beta$-functions of the 2HDM scalar potential parameters, which provide us with the
running values of these parameters with the renormalization scale. Since we are
not considering the fermion sector, we can write, for the most
general 2HDM~\cite{Haber:1993an,Branco:2011iw},\footnote{For simplicity of notation, we absorb the factors of
$16\pi^2$ in the definition of the $\beta$-functions.}
\begin{subequations}
\bea
\beta_{\lambda_1} &=&
12 \lambda_1^2 + 4 \lambda_3^2 + 4 \lambda_3 \lambda_4 + 2 \lambda_4^2
+ 2 \left| \lambda_5 \right|^2 + 24 \left| \lambda_6 \right|^2 \nonumber \\
 & &
 +\ \frac{3}{4}(3g^4 + g^{\prime 4} +2 g^2 g^{\prime 2}) -
 3\lambda_1 (3 g^2 + g^{\prime 2} ), \\
\beta_{\lambda_2} &=&
12 \lambda_2^2 + 4 \lambda_3^2 + 4 \lambda_3 \lambda_4 + 2 \lambda_4^2
+ 2 \left| \lambda_5 \right|^2 + 24 \left| \lambda_7 \right|^2 \nonumber \\
 & &
+\
\frac{3}{4}(3g^4 + g^{\prime 4} +2g^2 g^{\prime 2}) -3\lambda_2
(3g^2 +g^{\prime 2}),  \\
\beta_{\lambda_3}  &=&
\left( \lambda_1 + \lambda_2 \right) \left( 6 \lambda_3 + 2 \lambda_4 \right)
+ 4 \lambda_3^2 + 2 \lambda_4^2
+ 2 \left| \lambda_5 \right|^2
+ 4 \left( \left| \lambda_6 \right|^2 + \left| \lambda_7 \right|^2 \right)
+ 16\, \mathrm{Re} \left( \lambda_6 \lambda_7^\ast \right) \nonumber \\
 & &
+\ \frac{3}{4}(3g^4 + g^{\prime 4} -2g^2 g^{\prime 2}) - 3\lambda_3
(3g^2 +g^{\prime 2}), \\
\beta_{\lambda_4}  &=&
2 \left( \lambda_1 + \lambda_2 \right) \lambda_4
+ 8 \lambda_3 \lambda_4 + 4 \lambda_4^2
+ 8 \left| \lambda_5 \right|^2
+ 10 \left( \left| \lambda_6 \right|^2 + \left| \lambda_7 \right|^2 \right)
+ 4\, \mathrm{Re} \left( \lambda_6 \lambda_7^\ast \right) \nonumber \\
& &
+\ 3g^2 g^{\prime 2} - 3\lambda_4 (3g^2 +g^{\prime 2}), \\
\beta_{\lambda_5}  &=&
\left( 2 \lambda_1 + 2 \lambda_2 + 8 \lambda_3 + 12 \lambda_4 \right) \lambda_5
+ 10 \left( \lambda_6^2 + \lambda_7^2 \right) + 4 \lambda_6 \lambda_7
\nonumber \\
 & &
- \ 3\lambda_5 (3g^2 +g^{\prime 2}),  \\
\beta_{\lambda_6}  &=&
\left( 12 \lambda_1 + 6 \lambda_3 + 8 \lambda_4 \right) \lambda_6
+ \left( 6 \lambda_3 + 4 \lambda_4 \right) \lambda_7
+ 10 \lambda_5 \lambda_6^\ast + 2 \lambda_5 \lambda_7^\ast \nonumber \\
 & &
-\ 3\lambda_6 (3g^2 +g^{\prime 2}),  \\
\beta_{\lambda_7}  &=&
\left( 12 \lambda_2 + 6 \lambda_3 + 8 \lambda_4 \right) \lambda_7
+ \left( 6 \lambda_3 + 4 \lambda_4 \right) \lambda_6
+ 10 \lambda_5 \lambda_7^\ast + 2 \lambda_5 \lambda_6^\ast \nonumber \\
 & &
-\ 3\lambda_7 (3g^2 +g^{\prime 2}),
\label{eq:betal}
\eea
\end{subequations}
for the quartic couplings, and for the quadratic ones,
\begin{subequations}
\bea
\beta_{m_{11}^2} &=&
6 \lambda_1 m_{11}^2
+ \left( 4 \lambda_3 + 2 \lambda_4 \right) m_{22}^2
- 12\, \mathrm{Re} \left( m_{12}^2 \lambda_6^\ast \right),
\\
\beta_{m_{22}^2} &=&
\left( 4 \lambda_3 + 2 \lambda_4 \right) m_{11}^2
+ 6 \lambda_2 m_{22}^2
- 12\, \mathrm{Re} \left( m_{12}^2 \lambda_7^\ast \right),
\\
\beta_{m_{12}^2} &=&
- 6 \left( \lambda_6 m_{11}^2 + \lambda_7 m_{22}^2 \right)
+ \left( 2 \lambda_3 + 4 \lambda_4 \right) m_{12}^2
+ 6 \lambda_5 {m_{12}^2}^\ast.
\label{eq:betam}
\eea
\end{subequations}
In these relations we include the scalar and gauge couplings ($g$ and $g^\prime$ refer to the SU(2) and $\text{U(1)}_Y$ couplings), but not the Yukawa ones (though the analysis we will undertake here is easily extended to the fermionic sector as well).
The above $\beta$-functions allow us to verify whether the
relations obtained in the previous section among parameters are RGE invariant
to one-loop order.

Let us begin by giving some examples of symmetry relations which {\it are}
preserved by the one-loop $\beta$-functions:
\begin{itemize}
\item If the potential is invariant under the CP1 symmetry, there exists a basis where all
its parameters are real. It is then trivial to see from the expressions above that the RGE running
preserves $\Im\lambda_i = 0$, for any renormalization scale and any value of the parameters.
\item If the potential has a $Z_2$ symmetry, a basis exists for which $m^2_{12} = \lambda_6 =
\lambda_7 = 0$. Notice how these conditions then imply
\begin{equation}
\beta_{\lambda_6}  =
\beta_{\lambda_7}  =
\beta_{m_{12}^2} = 0,
\end{equation}
that is, the $m^2_{12} = \lambda_6 = \lambda_7 = 0$ ``point" in parameter space is a {\it fixed
point} for the RGE evolution of the parameters -- if the potential obeys those conditions at some
scale, it will obey them at any scale\footnote{On a side note, notice how one would still have
$\beta_{\lambda_6} = \beta_{\lambda_7} = 0$ even if $m_{12}^2\neq 0$. This occurs because a non-zero
value for the $m_{12}^2$ parameter yields a {\it soft breaking} of the $Z_2$ symmetry, not affecting
the renormalization of its dimensionless couplings.}.
\item With $\lambda_6 = \lambda_7 = 0$, notice that the $\beta$-function for $\lambda_5$ becomes
\beq
\beta_{\lambda_5}  =
\left[ 2 \lambda_1 + 2 \lambda_2 + 8 \lambda_3 + 12 \lambda_4
-  3 (3g^2 +g^{\prime 2}) \right] \lambda_5,
\eeq
which possesses a fixed point at $\lambda_5 = 0$ -- the conclusion is that the condition
$\lambda_5 = \lambda_6 = \lambda_7 = 0$ is also RGE invariant, and indeed we know it corresponds
to the quartic coupling conditions of the U(1) Peccei--Quinn  model.
\end{itemize}
Notice how this last example leads to one of the mass degeneracies required for the symmetry conditions
in the previous section. In fact, we had identified, in section~\ref{sec:cc} a possibility for $Z_2$
invariance which required two neutral scalars to be degenerate in mass, $M_2 = M_3$. Looking at
the mass matrix of Eq.~\eqref{eq:mmat}, this then implies $\lambda_5 = 0$, which leads us to a potential
with a symmetry larger than $Z_2$, namely the Peccei--Quinn model, with a continuous U(1) symmetry unbroken
by the vacuum.
We indeed found it, Case  BCC in section~\ref{sec:bcc}. This, then, is an example
whereupon mass degeneracy between scalars is preserved to all orders in perturbation theory.

One may follow the above procedure for all the parameter conditions presented in Table
5 of reference~\cite{Branco:2011iw}, and verify that all of those conditions are RGE invariant, at least
to one-loop order. We have
therefore a powerful tool to investigate the conditions we obtained for each case studied in the previous sections, and verify whether they are RGE stable. We provide several examples below
where that does {\it not} happen:
\begin{itemize}
\item In section~\ref{sec:cc}, we analysed Case \textoverline{BC}, wherein degeneracy of two neutral scalars
implied the following relation among couplings:
\begin{subequations}
\begin{equation} \label{Eq:BC-degeneracy}
 m_{22}^2 = (-2\lambda_1+\lambda_3+\lambda_4+\Re\lambda_5)v^2.
 \end{equation}
 Invoking the minimisation conditions relating $v$ to the parameters of the potential,
Eq.~\eqref{eq:m11}, we can rewrite this as
\begin{equation}
\lambda_1 \,m_{22}^2 = (-2\lambda_1+\lambda_3+\lambda_4+\Re\lambda_5) \,m_{11}^2.
\end{equation}
\end{subequations}
At this point, using the $\beta$-functions from Eqs.~\eqref{eq:betal}
and~\eqref{eq:betam}, it is very easy to confirm that
\beq
\beta(\lambda_1 \,m_{22}^2)\;\neq\;\beta \{(-2\lambda_1+\lambda_3+\lambda_4+\Re\lambda_5) \,m_{11}^2\}
\eeq
which implies that  one-loop corrections would destroy the equality \eqref{Eq:BC-degeneracy}.
Thus, already at the one-loop level, the parameter relation \eqref{Eq:BC-degeneracy}
corresponding to Case~\textoverline{BC} turns out to be merely an unstable tree-level fine-tuning of parameters that
does not hold in the perturbative expansion.
\item
In section~\ref{sec:bcc} we considered the Case \textoverline{ABBB} which requires a condition similar to Eq.~(\ref{Eq:BC-degeneracy}), but with $\Re\lambda_5=0$. This is also unstable, in analogy to the case above.
\item In section~\ref{sec:CP2} we studied the CP2 symmetry, which entails, in a given basis, $\lambda_1 =
\lambda_2$ and $\lambda_6 = \lambda_7 = \Im \lambda_5 = 0$.
Case \textoverline{BCD} further required (all couplings real)
\beq
\lambda_5 = 3\lambda_1 - \lambda_3 - \lambda_4.
\label{eq:5134}
\eeq
With the CP2 relations among the quartic couplings, we find (ignoring the gauge contributions
for the moment)
\bea
\beta_{\lambda_5} &=& 4\lambda_5 (\lambda_1 + 2\lambda_3 + 3\lambda_4)
\,,\nonumber \\
\beta_{3\lambda_1 - \lambda_3 - \lambda_4} &=
& 4\left[(9\lambda_1 - 3\lambda_3 -
2\lambda_4)\lambda_1 + (2\lambda_3 + \lambda_4)\lambda_3 -(3\lambda_1 - \lambda_3 - \lambda_4)^2 \right]\,,
\eea
and it is clear that the RGE running of the two sides of equation~\eqref{eq:5134}
are different, the relation is {\it not} preserved by radiative corrections. Including
the gauge contributions would not change this fact.
\item As a final example, in the SO(3) symmetry cases discussed in section~\ref{sect:section-SO3}, for which $\lambda_5=\lambda_6=\lambda_7 = 0$ and $\lambda_4 = \lambda_1 - \lambda_3$,
 mass degeneracies required $\lambda_1 = 0$. But for this symmetry class, the $\beta$-function for $\lambda_1$ is given by (ignoring the gauge contributions for the moment)
\beq
\beta_{\lambda_1} \,=\,
2 \lambda_3^2 ,
\eeq
which shows that $\lambda_1 = 0$ is not a fixed point in the RGE running. Again, including
the gauge contributions would not change this fact.
These are some of the examples of RGE instability found in the previous section. We leave the remainder as an exercise for the reader.
\end{itemize}

Some comments are here in order. Having shown that some of the cases discovered are in fact RGE unstable, we have stated that  under the running, we must remain within the symmetry class being considered, concluding that we will end up in one of the RGE-stable cases at loop level. This may seem counterintuitive at first, suggesting a discrete behavior under the running of the RGE. But as we pointed out in the discussion in Chapter \ref{Sec:analysis}, the RGE-unstable cases may be interpreted as special cases of the RGE-stable ones, meaning that they can be thought of as equivalent to the RGE-stable case with some fine-tuned mass degeneracy constraints imposed. At loop level the fine-tuned constraints will be lost, and we end up in the RGE-stable case.

We may illustrate this with a brief discussion of the conditions for CP1. Assume that at tree level we have imposed the (fine-tuned) mass degeneracy constraints $M_1=M_2=M_3$, {\em i.e} case \textoverline{A}. The full mass degeneracy implies that our potential (and vacuum) is CP1 invariant. Calculating one-loop corrections, however, the 
mass degeneracy will be lifted, resulting in an effective one-loop mass matrix that yields three neutral states with
masses {\em a priori} different. Then, 
after diagonalization of that matrix to obtain the one-loop physical eigenstates, we will obtain 
$e_k=q_k=0$ for some $k$ -- meaning, one neutral state that does not couple to gauge bosons or charged scalars, {\em i.e} the pseudoscalar state, expected in a CP1 model.
At tree level, though, with three mass-degenerate states, one can always construct a linear combination $H_k$ of those 
states such that it has vanishing couplings $e_k$ and $q_k$ (thus corresponding to case C). Running the RGE does not
change the fact that one of the states has $e_k=q_k=0$, but in the case of \textoverline{A} it simply means that
one will reach a given renormalization scale for which the (tree-level) relation $M_1=M_2=M_3$ holds. Case
\textoverline{A} can therefore be understood as a special case of case C, a particular point along the RGE trajectory.
The RGE unstable cases we have encountered, then, may be perceived as special points along the RGE-trajectories
of physical cases, wherein particular relations between tree-level parameters are found for specific values of
the renormalization scale -- but such relations do not hold at the one-loop level. 
A counterexample of such
behaviour is case BCC, wherein the mass degeneracy found between two neutral scalars will be
preserved by radiative corrections -- such degeneracy is ensured by a RGE-stable condition ($\lambda_5 = 0$), which is
not broken by the vacuum, thus it is preserved along all points on the RGE trajectory. The RGE-unstable 
cases discussed
in previous sections are therefore, as mentioned, special cases of the stable ones -- the extra conditions that 
characterize the RGE-unstable cases are the result of unphysical tree-level fine-tunings, and they ``migrate" to the 
RGE-invariant cases once radiative corrections are taken into account (see Fig.~\ref{fig:migration}).

\section{Discussion}
\label{Sec:summary}

We have seen that conditions for all six symmetries can be formulated in terms of constraints on physical quantities, masses and couplings. Each symmetry can be satisfied in a number of different ways, referred to as ``cases''. The recent results of Bento {\it et al} \cite{Bento:2020jei} translate into exactly the same set of ``cases'' as presented here, and are thus in full agreement with ours.
Some of the cases encountered, involving mass degeneracies, are unstable under radiative corrections. This result is to some extent surprising, as it turns out that sometimes,
i.e. for some ``cases'', even though the Lagrangian is symmetric under a given transformation, conditions that guarantee the invariance are not stable with respect to RGE.
The following cases remain stable at the one-loop level:
\begin{itemize}
\item[$\bullet$]{\bf CP1} [two constraints]
\begin{alignat}{2}
&\text{\bf Case C:} &\ \
&  e_k=q_k=0, \nonumber\\
&\text{\bf Case D:} &\ \
& 2D M_{H^\pm}^2=v^2
\left[
e_1q_1M_2^2M_3^2+e_2q_2M_1^2M_3^2+e_3q_3M_1^2M_2^2-M_1^2M_2^2M_3^2 \right], \nonumber\\
& & & \phantom{\!\!\bullet}
2D q=
(e_2q_3-e_3q_2)^2M_1^2+(e_3q_1-e_1q_3)^2M_2^2+(e_1q_2-e_2q_1)^2M_3^2+M_1^2M_2^2M_3^2,
\nonumber
\end{alignat}
Case C realizes an unbroken CP1 symmetry, whereas case D realizes a spontaneously broken CP1 symmetry.
\item[$\bullet$]$\bold{Z_2}$ [four constraints]
\begin{alignat}{2}
&\text{\bf Case CC:} &\ \
&e_j=q_j=e_k=q_k=0, \nonumber \\
&\text{\bf Case CD:} &\ \
&e_k=q_k=0, \quad 2(e_j^2M_i^2+e_i^2M_j^2)M_{H^\pm}^2=v^2(e_jq_jM_i^2+e_iq_iM_j^2-M_i^2M_j^2),\nonumber\\
& & & \phantom{\!\!\bullet}
2(e_j^2M_i^2+e_i^2M_j^2)q=(e_jq_i-e_iq_j)^2+M_i^2M_j^2. \nonumber
\end{alignat}
Case CC realizes an unbroken $Z_2$ symmetry, whereas case CD realizes a spontaneously broken $Z_2$ symmetry.
\item[$\bullet$]{\bf U(1)}  [five constraints]
\begin{alignat}{2}
&\text{\bf Case BCC:} &\ \
&M_j=M_k,\quad
e_j=q_j=e_k=q_k=0.\nonumber\\
&\text{\bf Case C$_0$D:} &\ \
&e_k=q_k=0, \quad 2(e_j^2M_i^2+e_i^2M_j^2)M_{H^\pm}^2=v^2(e_jq_jM_i^2+e_iq_iM_j^2-M_i^2M_j^2),\nonumber\\
& & & \phantom{\!\!\bullet}
2(e_j^2M_i^2+e_i^2M_j^2)q=(e_jq_i-e_iq_j)^2+M_i^2M_j^2,\quad M_k=0.\nonumber
\end{alignat}
Case BCC realizes an unbroken $U(1)$ symmetry, whereas case C$_0$D realizes a spontaneously broken $U(1)$ symmetry.
\item[$\bullet$]{\bf CP2} [six constraints]
\begin{alignat}{2}
&\text{\bf Case CCD:} &\ \
&e_j=q_j=e_k=q_k=0,\quad 2M_{H^\pm}^2=e_iq_i-M_i^2,\quad 2v^2q=M_i^2.\nonumber
\end{alignat}
Case CCD realizes a spontaneously broken CP2 symmetry. Note that the vacuum cannot be CP2 invariant.
\item[$\bullet$]{\bf CP3}  [seven constraints]
\begin{alignat}{2}
&\text{\bf Case BCCD:} &\ \
&e_j=q_j=e_k=q_k=0,\quad 2M_{H^\pm}^2=e_iq_i-M_i^2,\quad 2v^2q=M_i^2,\quad M_j=M_k,\nonumber\\
&\text{\bf Case C$_0$CD:} &\ \
&e_j=q_j=e_k=q_k=0,\quad 2M_{H^\pm}^2=e_iq_i-M_i^2,\quad 2v^2q=M_i^2,\quad M_j=0.\nonumber
\end{alignat}
Case BCCD realizes an unbroken CP3 symmetry, whereas case $C_0$CD realizes a spontaneously broken CP3 symmetry.
\item[$\bullet$]{\bf SO(3)} [eight constraints]
\begin{alignat}{2}
&\text{\bf Case B$_0$C$_0$C$_0$D:} &\
&M_j=M_k=0,\quad e_j=q_j=e_k=q_k=0,\nonumber\\
& & &2M_{H^\pm}^2=e_iq_i-M_i^2,\quad 2v^2q=M_i^2.\nonumber
\end{alignat}
Case B$_0$C$_0$C$_0$D realizes a spontaneously broken SO(3) symmetry. Note that the vacuum cannot be SO(3) invariant.
\end{itemize}

Let us comment on the importance of the constraints on $q$ and $\mchsq$ that come from imposing Case D. Remember that for the CP1, $Z_2$ and U(1) symmetries, the symmetry can either be unbroken or spontaneously broken by the vacuum. Note that the constraint D is absent whenever these symmetries are unbroken, whereas constraint D is always present if the symmetry is spontaneously broken. The symmetries CP2 and SO(3) cannot be unbroken, they must be spontaneously broken, and we see that the constraint D is present for these symmetries. Only CP3 remains to be discussed. It can either be unbroken or spontaneously broken, but constraint D is present in both these situations. To understand the presence of constraint D in the unbroken CP3 case, we recall that a CP3 invariant potential is also CP2 invariant. Thus, the unbroken CP3 will spontaneously break CP2, thereby explaining the presence of constraint D in the unbroken CP3 case.

It turns out that all the unstable cases are associated with some {\it mass degeneracy.}
However there exists three cases with {\it mass degeneracy} that are RGE stable
at the one-loop order, namely Cases BCC, BCCD and B$_0$C$_0$C$_0$D, where the degeneracy is directly implied by U(1), CP3 and SO(3) symmetries, respectively.
In this latter case the mass-degenerate pair is massless.
In all these three cases the mass-degenerate pair represents two states of opposite CP.

\begin{figure}[H]
	\centering
\includegraphics[height=11cm,angle=0]{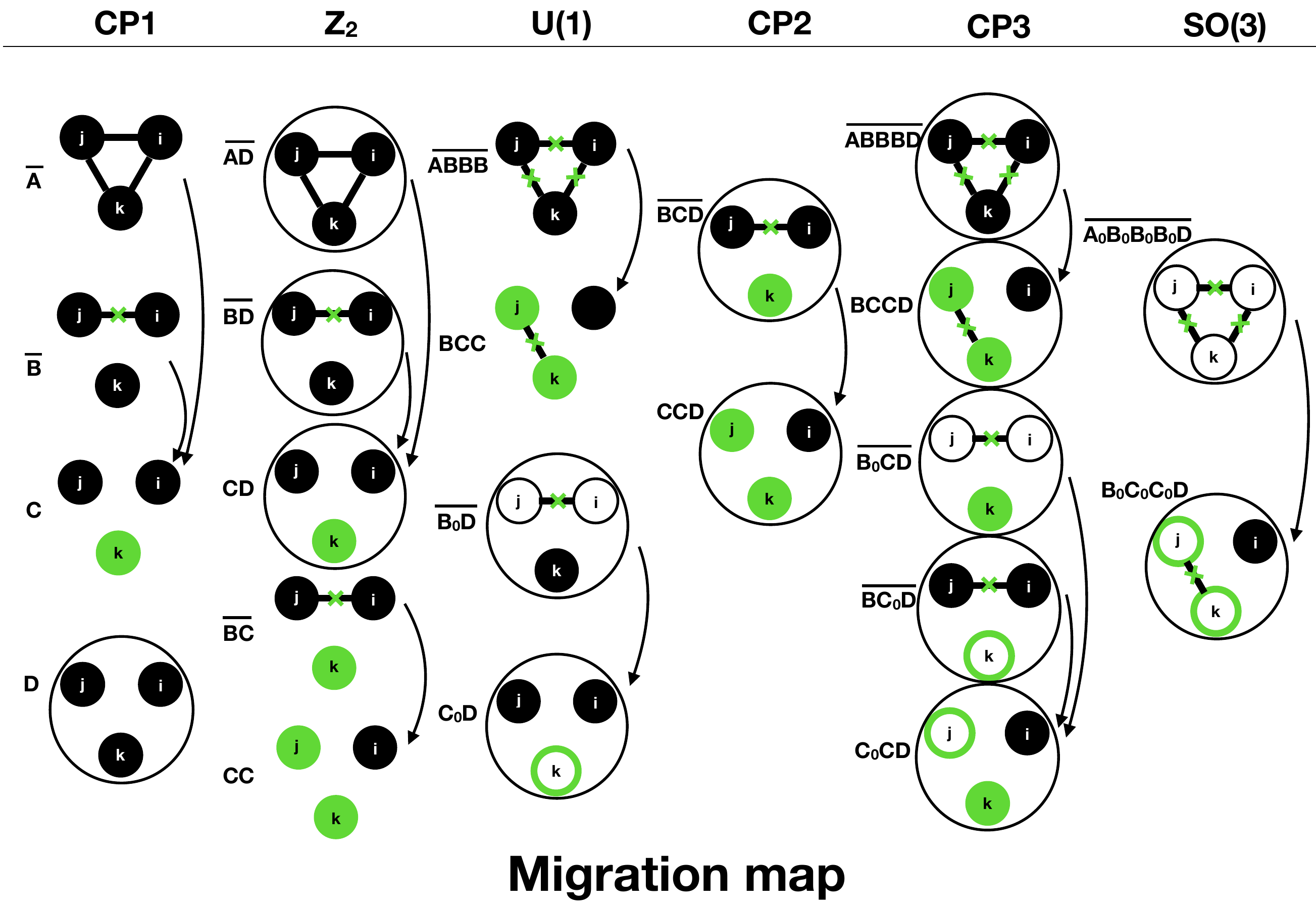}
	\caption{Migration of unstable cases under radiative corrections. Each of the three small circles labeled $i$, $j$ and $k$, represents a neutral scalar, with symbols as defined in section~\ref{sect:Results}. The cases \textoverline{ABBB} and \textoverline{ABBBD}, which both occur for two symmetries, are shown only under the higher symmetry. Cases enclosed by a larger circle have a constraint on $M_{H^\pm}$ and the quartic coupling $q$, whereas those without the ``overline'' are RGE stable at the one-loop order. Symbols in white refer to massless states.}
	\label{fig:migration}
\end{figure}

The way the unstable cases have ``migrated'' to stable ones is illustrated in Fig.~\ref{fig:migration}. Note that the stable cases have no ``bar'' over the identifier.

The different cases listed above all require a certain number of constraints. These numbers are collected in table~\ref{Table:numbers}, together with the complementary number, giving the number of free, independent physical parameters the model with a given symmetry contains.

\begin{table}[htb]
\begin{center}
\begin{tabular}{|c|c|c|c|c|c|c|c|c|}
\hline
\hline
& 2HDM& CP1 & $Z_2$ & U(1) & CP2 & CP3 & SO(3)\\
\hline
constraints& 0 & 2 & 4 & 5 & 6 & 7 & 8 \\
free parameters & 11  & 9 & 7 & 6 & 5 & 4 & 3\\
\hline
\end{tabular}
\end{center}
\caption{\label{Table:numbers} Starting with the unconstrained complex 2HDM, different symmetries impose a number of constraints, leaving the complementary number of free parameters.}
\end{table}

For example, a CP2-symmetric model has five independent physical parameters: $e_1=\pm v$, $q_1$, $M_1$, $M_2$ and $M_3$. The rest are either zero or given in terms of these: $e_2=e_3=0$, $q_2=q_3=0$, $M_{H^\pm}^2=\half(e_1q_1-M_1^2)$ and $q=M_1^2/(2v^2)$.

As another example, an SO(3)-symmetric model has three independent physical parameters: $e_1=\pm v$, $q_1$ and $M_1$. All other parameters of the model are either zero or given in terms of these three.

However, this does not mean that a model given in terms of {\it any} three parameters has SO(3) symmetry. It could also be a model with lower symmetry, but with parameters set to zero or related in a way which is unstable under radiative corrections.

We have performed a complete translation of conditions for 2HDM symmetries
in terms of physical, basis-invariant parameters. Our parameter set ${\cal P}$,
as detailed in eq.~\eqref{Eq:pcal}, includes the four scalar masses, the couplings of the three neutral
scalars to electroweak gauge bosons and to charged scalar pairs, and the quartic vertex interaction
among four charged fields, in a total of 11 physical parameters. All 2HDM symmetries leave less than 11 free parameters,
thus specific values for elements of ${\cal P}$, or relations between them, were found for each symmetry
class. We followed closely the work of ref.~\cite{Ferreira:2010yh}, wherein conditions for each of the
symmetries were found in terms of a bilinear formalism, which we then translated in terms of
physical parameters. We found that several of the symmetry conditions of~\cite{Ferreira:2010yh} implied
mass degeneracies among two or more scalars, which were then found to be unstable under renormalization.
Those specific symmetry cases corresponded to zero-measure regions of parameter space: tree-level
fine tunings corresponding to a very specific relation among couplings, not preserved by radiative
corrections. The ``migration map" just discussed shows how radiative corrections lift
mass degeneracies and lead those RGE unstable cases to symmetry cases stable under renormalization.

The remarkable benefit of expressing each 2HDM symmetry in terms of physical parameters is the
possibility of giving, in a simple and physically intuitive way, a description of each possible model.
For instance, at its simplest, the IDM, a 2HDM with a $Z_2$ symmetry left intact by spontaneous
symmetry breaking, is described in the simplest of fashions: a neutral scalar with SM-like couplings to gauge
bosons, two neutral scalars with no couplings to either electroweak gauge bosons nor to charged scalars. Other,
more elaborate conditions between physical parameters are found, and those arising from Case D (see
section~\ref{Sec:CP1} for the general conditions) are particularly interesting.
(See also Ref.~\cite{Grzadkowski:2016szj}.) Consider in fact the
conditions obtained for the model with a spontaneously broken $Z_2$ symmetry, eqs.~\eqref{BDes}
to~\eqref{Eq:Z2-spont-br}: the first specifies that one of the neutral states ($H_3$) does not couple to electroweak
gauge bosons or charged scalars -- {\em ergo}, it is the pseudoscalar $A$ (with mass $M_A$).
The formalism we used throughout this paper was necessary for a full discussion
of many different symmetry cases, but let us now use the ``usual" 2HDM notation and define
$\tan\beta = v_2/v_1$, let us assume $H_1$ is the lightest CP-even scalar, observed at the LHC (with mass
$M_h$), and therefore $H_2$ will be the heavier CP-even particle (with mass $M_H$). In order to obtain the ``usual" gauge couplings, we put $H_1=h$, $H_2=-H$, $H_3=A$ and introduce $\alpha=\tilde{\alpha}+\pi/2$
to get
\bea
&&e_h=e_1=v\sin(\beta-\tilde{\alpha}),\quad e_H=-e_2=v\cos(\beta-\tilde{\alpha}),\quad e_A=e_3=0\nonumber\\
&&q_h=q_1=0,\quad q_H=-q_2, \quad q_A=q_3=0,
\eea
where $\tilde{\alpha}$ is the ``usual" rotation angle of the CP-conserving 2HDM. Current LHC measurements indicate $\sin(\beta - \tilde{\alpha})\simeq 1$, also known as the alignment limit.
Much more fascinating is condition~\eqref{Eq:Z2-spont-br},
then rewritten as
\bea
M^2_{H^\pm} &=& \frac{v^2}{2}\,
\frac{e_H q_H M^2_h\,+\, e_h q_h M^2_H\,-\,M^2_h \,M^2_H}
{e_H^2 M^2_h \,+\, e_h^2 M^2_H}, \label{eq:mar1} \\
q &=& \frac{1}{2}\,
\frac{\left(e_H q_h - e_h q_H\right)^2\,+\,M^2_h M^2_H}
{e_H^2 M^2_h \,+\, e_h^2 M^2_H}\,. \label{eq:mar2}
\eea
Recalling that $q_h$ and $q_H$  are the couplings of the $h H^+ H^-$ and $H H^+ H^-$ vertices, whereas $q$ is the charged scalar quartic self interaction coupling, the first of these two
equations is truly remarkable. It gives us a relation between scalar masses and couplings which can
be used to put this model to the test -- or to predict the mass
of the charged scalar, for instance. Indeed, consider the (unrealistic, perhaps) possibility that
in some future collider a heavier CP-even state $H$ has been discovered, and its interactions to electroweak
gauge bosons measured with precision; in parallel, via measurements of its diphoton decays,
its coupling to an {\em undiscovered} charged scalar may be inferred indirectly from the data. Then,
eq.~\eqref{eq:mar1} would allow for a {\em prediction} of the charged scalar mass! And if that
prediction were then to be falsified by experimental data, it would be a clear signal that the
spontaneously broken $Z_2$ symmetric 2HDM was excluded. The second of these equations lets us in the same way predict the strength of the charged scalar self interaction coupling, which is assumed to be experimentally much more difficult to measure.
Further, considering that current LHC data point to
alignment ($e_h\to v$, $e_H\to 0$), eqs.~\eqref{eq:mar1} and \eqref{eq:mar2} give us predictions in
the alignment limit, to wit
\bea \label{Eq:alignment-1}
M^2_{H^\pm} &\simeq& \frac{1}{2}\,
\left(v q_h -\,M^2_h\right)\,,\\
q&\simeq&\frac{1}{2}\,\left(
\frac{q_H^2}{M_H^2}+\frac{M_h^2}{v^2}
\right).  \label{Eq:alignment-2}
\eea
Consider how the first of the above equations predicts the approximate value of the charged Higgs mass {\em using
quantities related to the lightest CP-even state, currently being measured at the LHC!} The second of these equations predicts the more inaccessible $q$ in terms of quantities related to both of the two CP-even scalars. In the alignment limit, there are also other couplings of the model that are related to $q$ (see Appendix C of \cite{Grzadkowski:2016szj}), and these are perhaps more accessible, i.e.
\bea
HHH^+H^-,\quad AAH^+H^-,\quad HHHH,\quad AAAA, \quad HHAA,
\eea
which are all proportional to $q$.

Identical expression to Eqs.~(\ref{Eq:alignment-1}) and (\ref{Eq:alignment-2}) are found for the case of a spontaneously broken U(1) symmetry (section \ref{sec:su1}).
In that case, if one were to gauge the U(1) group, the massless pseudoscalar would give mass to
a new $Z^\prime$ gauge boson, and the charged Higgs mass would be related to $M_h$ and $q_h$ by the approximate
expression given above, in the alignment limit.

This simple example shows, then, how powerful the formalism
developed in this work is, it allows us to cast symmetry conditions in terms of easy-to-understand
relations between measurable quantities, which can be put to experimental test and easily communicated.
 To say that a potential has a spontaneously broken $Z_2$ symmetry, one might
state: {\em in the basis where all couplings of the potential are real and
$m_{12}^2 = \lambda_6 = \lambda_7 = 0$ both doublets acquire non-zero vevs}. But the physics of that
case is much more interestingly described as: {\em a pseudoscalar exists, and two CP-even scalars have
couplings such that their relation to the charged mass is given by equation~\eqref{eq:mar1}}. With the
current work, such elegant and concise statements become possible, and 2HDM symmetries are
brought to light with newfound clarity.

In this work we only studied {\em exact} symmetries at the potential level, not considering the possibility
of soft breaking terms. Those certainly yield interesting phenomenologies, and enlarge the range of 2HDM
possibilities. For the $Z_2$ symmetry, for instance, the inclusion of a soft breaking term can make a
CP-breaking vacuum possible, or an explicitly CP-breaking potential to start with (the so-called
Complex 2HDM). The soft breaking terms will obviously change the relation between the physical parameter
set ${\cal P}$ and the parameters of the potential, though it will not enlarge the number of physical
parameters necessary to fully describe 2HDM symmetries. We will study softly broken 2HDM symmetries in a
forthcoming work.

\vspace*{10mm} {\bf Acknowledgements.}
P.M.F.'s research is supported
by \textit{Funda\c c\~ao para a Ci\^encia e a Tecnologia} (FCT)
through contracts
UIDB/00618/2020, UIDP/00618/2020, CERN/FIS-PAR/0004/2019, CERN/FIS-PAR/0014/2019
and by HARMONIA project's contract UMO-2015/18/M/ST2/00518.
The work of B.G. is supported in part by the National Science Centre (Poland) as a research project, decision no 2017/25/B/ST2/00191
and HARMONIA project under contract no. UMO-2015/18/M/ST2/00518 (2016{2019).
The research of P.O. has been supported in part by the Research Council of Norway.


\appendix
\section{Translating parameters of the potential}
\label{Translation}
\setcounter{equation}{0}
A method of translating the parameters of the potential in the Higgs basis into physical parameters has been explained in detail in \cite{Ogreid:2018bjq}. We quote here the explicit relations used:
\begin{eqnarray}
m_{11}^2&=&\frac{e_1^2 M_1^2+e_2^2 M_2^2+e_3^2 M_3^2}{v^2},\\
m_{12}^2&=&\frac{e_1f_1 M_1^2+e_2f_2 M_2^2+e_3f_3 M_3^2}{v^2},\\
m_{22}^2&=&-2M_{H^\pm}^2+e_1q_1+e_2q_2+e_3q_3,\\
\lambda_1&=&\frac{e_1^2 M_1^2+e_2^2 M_2^2+e_3^2 M_3^2}{v^4},\\
\lambda_2&=&2q,\\
\lambda_3&=&\frac{e_1q_1+e_2q_2+e_3q_3}{v^2},\\
\lambda_4&=&-\frac{2 M_{H^\pm}^2}{v^2}-\frac{e_1^2 M_1^2+e_2^2 M_2^2+e_3^2 M_3^2}{v^4}+\frac{ M_1^2+ M_2^2+ M_3^2}{v^2},\\
\lambda_5&=&\frac{f_1^2 M_1^2+f_2^2 M_2^2+f_3^2 M_3^2}{v^4},\\
\lambda_6&=&\frac{e_1f_1 M_1^2+e_2f_2 M_2^2+e_3f_3 M_3^2}{v^4},\\
\lambda_7&=&\frac{f_1q_1+f_2q_2+f_3q_3}{v^2}.
\end{eqnarray}

\label{Sec:Eigensystems_Appendix}
\setcounter{equation}{0}
Before discussing eigenvalues of the matrix $E$ it is useful to  introduce the following notation  \cite{Grzadkowski:2016szj}:
\begin{align} \label{Eq:d_ijk}
d_{ijk}&=\frac{q_{1}^i M_1^{2j}e_1^k+q_{2}^i M_2^{2j}e_2^k+q_{3}^i M_3^{2j}e_3^k}{v^{i+2j+k}},\\
m_+&=\frac{M_{H^\pm}^2}{v^2}.
\end{align}
Then, the characteristic polynomial of the matrix $E$ can be written as
\bea
a\Lambda^3+b\Lambda^2+c\Lambda+d=0,
\eea
where
\bea
a&=&-1,\\
b&=&\frac{1}{8} \left(4 d_{010}-3 d_{012}-2 d_{101}-8 m_++2 q\right),\\
c&=&\frac{1}{16} \left[m_+ \left(4 d_{010}-2 d_{012}-4 d_{101}+4 q\right)+d_{010} \left(3 d_{012}+2 d_{101}-2 q\right)\right.\nonumber\\
&&\left.\hspace*{1cm}-2 d_{010}^2+2 q d_{012}+2 d_{020}-3 d_{022}-d_{101}^2-2 d_{111}+d_{200}-4 m_+^2\right],\\
d&=&\frac{1}{64} \bigl\{-2 d_{010} \left(m_+ d_{012}+2 q d_{012}+d_{022}-2 m_+ d_{101}-d_{101}^2-2 d_{111}+d_{200}+2 m_+ q\right) \nonumber\\
&&\hspace*{1cm}+d_{010}^2 \left(d_{012}-2 d_{101}+2 q\right)+d_{012} \left[2 \left(d_{200}+2 m_+ q+m_+^2\right)-d_{020}\right]\nonumber\\
&&\hspace*{1cm}+2 \left(d_{020} d_{101}-q d_{020}+m_+ d_{022}+2 q d_{022}+d_{032}-2 d_{101} d_{111}\right.\\
&&\hspace*{2cm}\left.-2 m_+^2 d_{101}-m_+ d_{101}^2-2 m_+ d_{111}-2 d_{121}+m_+ d_{200}+d_{210}+2 m_+^2 q\right)\bigr\}.\nonumber
\eea
The discriminants determining the number of repeated eigenvalues are given by
\bea
\Delta&=&18abcd-4b^3d+b^2c^2-4ac^3-27a^2d^2,\\
\Delta_0&=&b^2-3ac.
\eea
Also, the following abbreviation is convenient for writing out the eigenvalues,
\begin{equation}
\Delta_1=2b^3-9abc+27a^2d.
\end{equation}
Whenever $\Delta=0$, there are degenerate eigenvalues. If in addition $\Delta_0\neq0$, the eigenvalues are doubly degenerate, and if $\Delta_0=0$ they are triply degenerate.
For later use, we note that we can write the second one of these discriminants as
\bea
\Delta_0&=&\frac{1}{64v^8}\left[
v^8\left(2q-2d_{101}-2d_{010}+3d_{012}+4m_+\right)^2\right.\nonumber\\
&&\hspace*{1.3cm}+12v^4\left(e_1q_2-e_2q_1-\frac{e_1e_2}{v^2}(M_2^2-M_1^2)\right)^2\nonumber\\
&&\hspace*{1.3cm}+12v^4\left(e_1q_3-e_3q_1-\frac{e_1e_3}{v^2}(M_3^2-M_1^2)\right)^2\nonumber\\
&&\hspace*{1.3cm}+12v^4\left(e_2q_3-e_3q_2-\frac{e_2e_3}{v^2}(M_3^2-M_2^2)\right)^2\nonumber\\
&&\left.\hspace*{1.3cm}
+12\lambda\left(e_1^2(M_3^2-M_2^2),-e_2^2(M_3^2-M_1^2),e_3^2(M_2^2-M_1^2)\right)
\right]\label{delta0alt},
\eea
where $\lambda(x,y,z)$ is the K\"all\'en function defined in Appendix \ref{Sec:Kallen}.
This form is useful to determine if and when this discriminant vanishes.

The three eigenvalues of $E$ are in general given as
\bea
\Lambda_m&=&\frac{1}{3}\left(
b+\zeta^{m-1}C+\frac{\Delta_0}{\zeta^{m-1}C}\label{generaleigenvalues}
\right),
\eea
where $m=1,2,3$, $\zeta=-\half+\half\sqrt{3}\,i$ and $C=\sqrt[3]{\frac{\Delta_1+\sqrt{\Delta_1^2-4\Delta_0^3}}{2}}$ (provided $\Delta_0\neq0$)\footnote{In case $\Delta_0=0$, we may need to choose a different sign in front of the square root, i.e. if our expression yields $C=0$, then we need to replace our expression for $C$ with $C=\sqrt[3]{\frac{\Delta_1-\sqrt{\Delta_1^2-4\Delta_0^3}}{2}}$ in order to get the correct eigenvalues.}.

The eigenvectors of $E$ can in the general case be written as (note that they are not normalized):
\bea
\vec{e}_m&=&\frac{1}{2v^6}
\left[(e_1M_1^2-v^2q_1)(2v^2\Lambda_m+\mchsq)+e_1M_2^2M_3^2\right]\vec{F}_1^a\nonumber\\
&&+\frac{1}{2v^6}
\left[(e_2M_2^2-v^2q_2)(2v^2\Lambda_m+\mchsq)+e_2M_1^2M_3^2\right]\vec{F}_2^a\nonumber\\
&&+\frac{1}{2v^6}
\left[(e_3M_3^2-v^2q_3)(2v^2\Lambda_m+\mchsq)+e_3M_1^2M_2^2\right]\vec{F}_3^a\nonumber\\
&&+\frac{1}{2v^5}M_1^2(e_3q_2-e_2q_3)\vec{F}_1^b\nonumber\\
&&+\frac{1}{2v^5}M_2^2(e_1q_3-e_3q_1)\vec{F}_2^b\nonumber\\
&&+\frac{1}{2v^5}M_3^2(e_2q_1-e_1q_2)\vec{F}_3^b\nonumber\\
&&+\frac{1}{v}\left[8\Lambda_m^2+4(d_{012}-d_{010}+2m_+)\Lambda_m\right.\nonumber\\
&&\hspace*{0.8cm}\left.+2m_+^2+2m_+(d_{012}-d_{010})+d_{010}^2-2d_{010}d_{012}-d_{020}+2d_{022}\right]\vec{F}^c.\label{generaleigenvectors}
\eea
It is worth noting here, that the zero vector is by definition not an eigenvector of a matrix. We will encounter physical configurations where the above expression for $\vec{e}_m$ reduces to $\vec{0}$. For those configurations we will need to work out the eigenvectors anew. In particular, this is necessary for Configurations 1--3 listed in section~\ref{Sec:Vanishing_imJ1}.

\subsection{Configuration 1}
The three eigenvalues are in this physical configuration (see section~\ref{Sec:Vanishing_imJ1}) given as
\bea
\Lambda_1&=&\frac{M_1^2-M_{H^\pm}^2}{2v^2},\\
\Lambda_{2,3}&=&\frac{1}{16v^2}\bigg[5M_1^2-4M_{H^\pm}^2-2(e_1q_1+e_2q_2+e_3q_3)+2v^2q\\
&&\hspace*{1.2cm}\pm\sqrt{[3M_1^2-4M_{H^\pm}^2+2(e_1q_1+e_2q_2+e_3q_3)-2v^2q]^2+16{\cal Q}^2}\bigg].\nonumber
\eea
For the eigenvalue $\Lambda_1$, our general expression (\ref{generaleigenvectors}) for the eigenvectors reduces to $\vec{0}$, so for this eigenvalue we need to calculate the associated eigenvector anew. For the two remaining eigenvalues, we can use the general expression, and after some simplifications we arrive at
\bea
\vec{e}_1&=&\frac{1}{v^2}\left(q_1\vec{F}_1^b+q_2\vec{F}_2^b+q_3\vec{F}_3^b\right),\label{case1alphaeigen1}\\
\vec{e}_{2,3}&=&\frac{1}{v^2}\left(q_1\vec{F}_1^a+q_2\vec{F}_2^a+q_3\vec{F}_3^a\right)+\frac{4}{v^3}(M_1^2-M_{H^\pm}^2-2v^2\Lambda_{2,3})\vec{F}^c.\label{case1alphaeigen3}
\eea
provided ${\cal Q}^2\neq0$, in which case they all reduce to $\vec{0}$. Thus, the situation when ${\cal Q}^2=0$ needs separate treatment. The eigenvalues are then  given by
\bea
\Lambda_{1,2}&=&\frac{M_1^2-M_{H^\pm}^2}{2v^2},\\
\Lambda_{3}&=&\frac{1}{8v^2}(M_1^2-2(e_1q_1+e_2q_2+e_3q_3)+2v^2q),
\eea
with corresponding eigenvectors
\bea
\vec{e}_1&=&(1,0,0),\label{case1betaeigen1}\\
\vec{e}_2&=&(0,1,0),\\
\vec{e}_3&=&(0,0,1)\label{case1betaeigen3}.
\eea
Thus, it is convenient to divide Configuration 1 into two different sub-configurations, and refer to these sub-configurations when discussing the different symmetries.
\begin{alignat}{2}
&\!\!\bullet &\ \
&\text{\bf Configuration 1$\alpha$: }  M_1=M_2=M_3,\, {\cal Q}^2\neq0. \nonumber \\
&\!\!\bullet &\ \
&\text{\bf Configuration 1$\beta$: }  M_1=M_2=M_3,\, {\cal Q}^2=0. \nonumber
\end{alignat}
\subsection{Configuration 2}
The three eigenvalues of the $E$-matrix are in this physical configuration given as
\bea
\Lambda_1&=&\frac{M_i^2-M_{H^\pm}^2}{2v^2},\\
\Lambda_{2,3}&=&\frac{1}{16v^4}\left\{2v^2[2M_i^2+2M_k^2-2M_{H^\pm}^2-(e_1q_1+e_2q_2+e_3q_3)+v^2q]\right.\nonumber\\
&&\hspace*{1.2cm}-3\left[(e_i^2+e_j^2)M_i^2+e_k^2M_k^2\right]\nonumber\\
&&\left.\hspace*{1.2cm}\pm\sqrt{
\begin{aligned}
  &\left[2v^2(2M_i^2+2M_k^2-2M_{H^\pm}^2+e_1q_1+e_2q_2+e_3q_3-v^2q)\right.\\
  &\left. \phantom{xxxxxxxxxxxxxxxxxxxxx}-5\left((e_i^2+e_j^2)M_i^2+e_k^2M_k^2\right)\right]^2\\
	&+16\left[v^2(e_iq_k-e_kq_i)+e_ie_k(M_i^2-M_k^2)\right]^2\\
	&+16\left[v^2(e_jq_k-e_kq_j)+e_je_k(M_i^2-M_k^2)\right]^2
\end{aligned}
}
\right\},\nonumber\\
\eea
subject to the constraint $e_jq_i-e_iq_j=0$.
The eigenvalue $\Lambda_1$ inserted into our general expression (\ref{generaleigenvectors}) for the eigenvectors makes $\vec{e}_1$ reduce to $\vec{0}$, so for this eigenvalue we need to calculate the associated eigenvector anew. We find
\bea
\vec{e}_1&=&\frac{1}{v}\vec{F}_k^b,\label{case2alphaeigen1}\\
\vec{e}_{2,3}&=&\frac{1}{v^6}\left\{e_i\left[v^2(e_iq_k-e_kq_i)+e_ie_k(M_i^2-M_k^2)\right]\right.\nonumber\\
&&\left.\hspace*{0.8cm}+e_j\left[v^2(e_jq_k-e_kq_j)+e_je_k(M_i^2-M_k^2)\right]\right\}\vec{F}_k^a\nonumber\\
&&+\frac{4(e_i^2+e_j^2)}{v^7}\left[e_k^2(M_i^2-M_k^2)+v^2(M_k^2-M_{H^\pm}^2-2v^2\Lambda_{2,3})\right]\vec{F}^c,\label{case2alphaeigen3}
\eea
provided $\left[v^2(e_iq_k-e_kq_i)+e_ie_k(M_i^2-M_k^2)\right]^2
+\left[v^2(e_jq_k-e_kq_j)+e_je_k(M_i^2-M_k^2)\right]^2\neq0$ and $e_i^2+e_j^2\neq0$.

If $v^2(e_iq_k-e_kq_i)+e_ie_k(M_i^2-M_k^2)=0$ and $v^2(e_jq_k-e_kq_j)+e_je_k(M_i^2-M_k^2)=0$, then $\vec{e}_2$ or $\vec{e}_3$ reduces to $\vec{0}$, and that eigenvector must be calculated anew. The eigenvalues simplify to
\bea
\Lambda_1&=&\frac{M_i^2-M_{H^\pm}^2}{2v^2},\\
\Lambda_2&=&\frac{e_k^2M_i^2+(e_i^2+e_j^2)M_k^2-v^2M_{H^\pm}^2}{2v^4},\\
\Lambda_3&=&\frac{2v^4q+(e_i^2+e_j^2)M_i^2+e_k^2M_k^2-2v^2(e_iq_i+e_jq_j+e_kq_k)}{8v^4},
\eea
and the eigenvectors are then given as
\bea
\vec{e}_1&=&\frac{1}{v}\vec{F}_k^b,\label{case2betaeigen1}\\
\vec{e}_2&=&\frac{1}{v}\vec{F}_k^a,\\
\vec{e}_3&=&\frac{1}{v}\vec{F}^c,\label{case2betaeigen3}
\eea
provided $e_i^2+e_j^2\neq0$.

Whenever $e_i=e_j=0$, all eigenvectors of (\ref{case2alphaeigen1})--(\ref{case2alphaeigen3}) reduce to $\vec{0}$ and must be calculated anew. In this situation the eigenvectors are given by
\bea
\vec{e}_1&=&\frac{1}{v^2}\left(q_i\vec{F}_i^b+q_j\vec{F}_j^b\right),\label{case2gammaeigen1}\\
\vec{e}_{2,3}&=&\frac{1}{v^2}(q_i\vec{F}_i^a+q_j\vec{F}_j^a)+\frac{4}{v^3}(M_i^2-\mchsq-2v^2\Lambda_{2,3})\vec{F}^c,\label{case2gammaeigen3}
\eea
provided $q_i^2+q_j^2\neq0$. Finally, if $e_i=e_j=q_i=q_j=0$, also these eigenvectors reduce to $\vec{0}$, and must be calculated anew. The eigenvalues are then given by
\bea
\Lambda_{1,2}&=&\frac{M_i^2-M_{H^\pm}^2}{2v^2},\\
\Lambda_{3}&=&\frac{1}{8v^2}(M_k^2-2e_kq_k+2v^2q),
\eea
with corresponding eigenvectors
\bea
\vec{e}_1&=&(1,0,0),\label{case2deltaeigen1}\\
\vec{e}_2&=&(0,1,0),\\
\vec{e}_3&=&(0,0,1).\label{case2deltaeigen3}
\eea
Then, it will be convenient to divide all these different situations into four different sub-configurations, and refer to these sub-configurations when discussing the different symmetries:
\begin{alignat}{2}
&\!\!\bullet &\ \
&\text{\bf Configuration 2$\alpha$: }  M_i=M_j,\, e_jq_i-e_iq_j=0,\,
e_i^2+e_j^2\neq0,
\nonumber\\
& & &\phantom{\text{\bf Configuration 2$\alpha$: }} \left[v^2(e_iq_k-e_kq_i)+e_ie_k(M_i^2-M_k^2)\right]^2\nonumber\\
& & &\phantom{\text{\bf Configuration 2$\alpha$: }}
+\left[v^2(e_jq_k-e_kq_j)+e_je_k(M_i^2-M_k^2)\right]^2\neq0.\nonumber\\
&\!\!\bullet &\ \
&\text{\bf Configuration 2$\beta$: }  M_i=M_j,\, e_jq_i-e_iq_j=0,\,
e_i^2+e_j^2\neq0,
\nonumber\\
& & &\phantom{\text{\bf Configuration 2$\beta$: }} v^2(e_iq_k-e_kq_i)+e_ie_k(M_i^2-M_k^2)=0,\nonumber\\
& & &\phantom{\text{\bf Configuration 2$\beta$: }}
v^2(e_jq_k-e_kq_j)+e_je_k(M_i^2-M_k^2)=0.\nonumber\\
&\!\!\bullet &\ \
&\text{\bf Configuration 2$\epsilon$: }  M_i=M_j,\,e_i=e_j=0,\, q_i^2+q_j^2\neq0. \nonumber \\
&\!\!\bullet &\ \
&\text{\bf Configuration 2$\zeta$: }  M_i=M_j,\,e_i=e_j=q_i=q_j=0. \nonumber
\end{alignat}
\subsection{Configuration 3}
The three eigenvalues are in this physical configuration given as
\bea
\Lambda_1&=&\frac{M_k^2-M_{H^\pm}^2}{2v^2},\\
\Lambda_{2,3}&=&\frac{1}{16v^4}
\left\{
2v^2[2M_i^2+2M_j^2-2M_{H^\pm}^2-(e_iq_i+e_jq_j)+v^2q]-3(e_i^2M_i^2+e_j^2M_j^2)\right.\\
&&\left.\pm\sqrt{
\splitfrac{\left[2v^2(2M_i^2+2M_j^2-2M_{H^\pm}^2+e_iq_i+e_jq_j-v^2q)-5(e_i^2M_i^2+e_j^2M_j^2)\right]^2}{+16\left[v^2(e_iq_j-e_jq_i)+e_ie_j(M_i^2-M_j^2)\right]^2}
}
\right\},\nonumber
\eea
For Configuration 3, the eigenvalue $\Lambda_1$ inserted into our general expression (\ref{generaleigenvectors}) for the eigenvectors makes $\vec{e}_1$ reduce to $\vec{0}$, so for this eigenvalue we need to calculate the associated eigenvector anew. For the two remaining eigenvalues, we can use the general expression, and after some simplifications we arrive at
\bea
\vec{e}_1&=&\frac{1}{v}\vec{F}_k^a,\label{case3alphaeigen1}\\
\vec{e}_{2,3}&=&-\frac{1}{v^5}\left[v^2(e_iq_j-e_jq_i)+e_ie_j(M_i^2-M_j^2)\right]\vec{F}_k^b\nonumber\\
&&+\frac{4}{v^5}(e_j^2M_i^2+e_i^2M_j^2-v^2M_{H^\pm}^2-2v^4\Lambda_{2,3})\vec{F}^c,\label{case3alphaeigen3}
\eea
provided $v^2(e_iq_j-e_jq_i)+e_ie_j(M_i^2-M_j^2)\neq0$.

Whenever $v^2(e_iq_j-e_jq_i)+e_ie_j(M_i^2-M_j^2)=0$,  then $\vec{e}_{2}$ or $\vec{e}_3$ reduces to $\vec{0}$, and that eigenvector must be calculated anew. The eigenvalues simplify to
\bea
\Lambda_1&=&\frac{M_k^2-M_{H^\pm}^2}{2v^2},\\
\Lambda_2&=&\frac{e_j^2M_i^2+e_i^2M_j^2-v^2M_{H^\pm}^2}{2v^4},\\
\Lambda_3&=&\frac{2v^4q+e_i^2M_i^2+e_j^2M_j^2-2v^2(e_iq_i+e_jq_j)}{8v^4},
\eea
and the eigenvectors are then given as
\bea
\vec{e}_1&=&\frac{1}{v}\vec{F}_k^a,\label{case3betaeigen1}\\
\vec{e}_2&=&\frac{1}{v}\vec{F}_k^b,\\
\vec{e}_3&=&\frac{1}{v}\vec{F}^c.\label{case3betaeigen3}
\eea
Then, it will be convenient to divide these different situations into two different sub-configurations, and refer to these sub-configurations when discussing the different symmetries.
\begin{alignat}{2}
&\!\!\bullet &\ \
&\text{\bf Configuration 3$\alpha$: }  e_k=q_k=0,\,  v^2(e_iq_j-e_jq_i)+e_ie_j(M_i^2-M_j^2)\neq0. \nonumber \\
&\!\!\bullet &\ \
&\text{\bf Configuration 3$\beta$: }  e_k=q_k=0,\,  v^2(e_iq_j-e_jq_i)+e_ie_j(M_i^2-M_j^2)=0. \nonumber
\end{alignat}
\section{The K\"all\'en function}
\label{Sec:Kallen}
\setcounter{equation}{0}
The K\"all\'en function is defined as
\bea
\lambda(x,y,z)=x^2+y^2+z^2-2xy-2xz-2yz.
\eea
It is symmetric under interchange of two of its arguments, and also symmetric under a simultaneous change of sign of all three arguments,
\bea
\lambda(-x,-y,-z)=\lambda(x,y,z).
\eea
The K\"all\'en function is always positive whenever one argument has opposite sign of the other two.
Assuming $a,b,c$ real, we see this from
\bea
\lambda(a^2,-b^2,c^2)=\lambda(-a^2,b^2,-c^2)=((a-c)^2+b^2)((a+c)^2+b^2).
\eea

\bibliography{biblio}
\bibliographystyle{JHEP}

\end{document}